\documentclass[aps,pre,onecolumn,longbibliography,superscriptaddress,floatfix,nofootinbib,showkeys]{revtex4-2}
\usepackage{amsfonts}
\usepackage{amssymb}
\usepackage{amsmath}
\usepackage{graphicx}
\usepackage{url}
\usepackage{times}

\usepackage[normalem]{ulem}
\usepackage{xcolor}

\begin{document}

\title{Vaccination compartmental epidemiological models for the delta and omicron
SARS-CoV-2 variants}

\author{J. Cuevas-Maraver}

\affiliation{Grupo de F\'{\i}sica No Lineal, Departamento de F\'{\i}sica Aplicada I,
Universidad de Sevilla. Escuela Polit\'{e}cnica Superior, C/ Virgen de \'{A}frica, 7, 41011-Sevilla, Spain}
\affiliation{Instituto de Matem\'{a}ticas de la Universidad de Sevilla (IMUS). Edificio
Celestino Mutis. Avda. Reina Mercedes s/n, 41012-Sevilla, Spain}

\author{P.G. Kevrekidis}
\author{Q.Y. Chen}
\affiliation{Department of Mathematics and Statistics, University of Massachusetts Amherst,
Amherst, MA 01003, USA}

\author{G.A. Kevrekidis}
\affiliation{Department of Applied Mathematics and Statistics, Johns Hopkins University, Baltimore, MD 21218, USA}
\affiliation{Los Alamos National Laboratory, Los Alamos NM, USA}
\affiliation{Mathematical Institute for Data Science, Johns Hopkins University, Baltimore MD, USA}

\author{Y. Drossinos}
\affiliation{Thermal Hydraulics \& Multiphase Flow Laboratory,
Institute of Nuclear \& Radiological Sciences and Technology, Energy \& Safety,  \\
N.C.S.R. ``Demokritos",  GR-15341 Agia Paraskevi, Greece}

\begin{abstract}
    We explore the inclusion of
    vaccination in compartmental epidemiological
    models concerning the delta and
    omicron variants of the SARS-CoV-2 virus that
    caused the COVID-19 pandemic. We expand
    on our earlier compartmental-model
    work by incorporating vaccinated populations. We present
    two classes of models that differ depending on the immunological
    properties of the variant. The first one is for the delta variant,
    where we do not follow the dynamics of the vaccinated
    individuals since infections of vaccinated individuals were rare. The
    second one for the far more
    contagious omicron variant incorporates the evolution
    of the infections within the vaccinated cohort.
    We explore comparisons with available data involving
    two possible classes of counts, fatalities and hospitalizations.
    We present our results for two regions,
    Andalusia and Switzerland (including the Principality of Liechtenstein),
     where the necessary data are available. In the majority of the considered
    cases, the models are found to yield
    good agreement with the data and have a reasonable predictive
    capability beyond their training window, rendering them
    potentially useful tools for the interpretation
    of the COVID-19 and further pandemic waves, and for the
    design of intervention strategies during these waves.
\end{abstract}

\date{\today}

\keywords{epidemiological model, SARS-CoV-2, vaccination, delta variant, omicron variant}

\maketitle

\section{Introduction}
\label{sec:Introduction}

Over the last two and a half years, the COVID-19 pandemic
has been deemed responsible, to date, for 760 million confirmed cases,
and over 6.8 million deaths worldwide, as of this writing
and according to the World Health Organization COVID-19 dashboard.
As such, its emergence wreaked havoc in life as we knew it
throughout the world and forced a dramatic modification of
our social and economic activities during this interval.
At the same time, it triggered a global mobilization of the
scientific community to produce vaccines rapidly,
especially through (thankfully, by that time, fairly mature)
technology of mRNA-based methods. This effort
led to the remarkable result of having a vaccine against
SARS-CoV-2 within a year of its emergence. Nevertheless,
this was far from the end of the story, as new variants
of the SARS-CoV-2 virus kept emerging within 2020 and 2021.
The so-called delta variant appeared
in India in late 2020 and it had spread to 179 countries
by November 2021. Subsequently, the delta variant was superseded
by the so-called omicron variant that was reported in
South Africa on November 2021, and subsequently became
rapidly the predominant variant of SARS-CoV-2 thereafter.

The theoretical and mathematical modeling of infectious
diseases such as COVID-19 has a long and time-honored history
since the classic work of Kermack and McKendrick~\cite{kermack}.
Moreover, relevant efforts have been summarized in numerous
venues in recent years, such as, e.g.,~\cite{hethcote,castillo2011,chen2014modeling},
to mention only a few. The urgency and severity of the COVID-19
pandemic brought about an intense effort on the side of the
mathematical and physical communities to develop analytical models
and computational tools that could be used to examine the
unprecedented volume of available data regarding the temporal
(and spatial)
evolution of the pandemic and to make predictions for the
weeks (or in some cases month(s)) ahead. A notable example
of comparison of such efforts can be seen in, e.g.,
websites such as~\cite{ForeCast}.
Relevant modeling efforts have now been summarized
in a number of reviews such as~\cite{cao21,shakeel},
including ones of specialized modeling aspects such
as the study of metapopulation network models~\cite{review_meta},
while other works summarized the challenges and difficulties
of associated modeling~\cite{bertozzi2020,holmdahl2020}.

Over the last year, a large portion of the focus of the modeling
efforts has shifted towards the inclusion of vaccination in epidemniological models.
While a lot of information is available regarding
the effectiveness and efficacy of vaccines~\cite{tregoning2021}
(see also websites such as \cite{Vaccines})
mathematical models can still be quite useful in a number of
ways, including in guiding and informing distribution strategies
thereof~\cite{wagner2022,StilianakisVaccine2022}. It is in that light that numerous
compartmental epidemiological models with vaccination strategies have
arisen in the literature~\cite{marinov2022,vacc2021},
including some specific to different
geographical locations~\cite{MACINTYRE20222506}
and to different social infrastructures, such as
nursing homes~\cite{10.1093/cid/ciab517}.
While the relevant models feature different levels of complexity
starting from SIRV (Susceptible-Infected-Recovered-Vaccinated)
extensions of the classic SIR (Susceptible-Infected-Recovered)~\cite{marinov2022}
model and progressively extending to multicomponent models such
as, e.g.,~\cite{rychtar2021}, our aim here is to
build systematically on the earlier modeling attempt of~\cite{cuevas2021}
by considering SARS-CoV-2 variants that affect differently the vaccinated
population.

More concretely, our aim is to present a model of the
omicron variant (model A and its two implementations A1 and A2),
in which its highly contagious nature
allows for so-called breakthrough infections,
whereby vaccinated individuals may still be infected.
In that light, we account for the standard populations
of our earlier work~\cite{cuevas2021}, including exposed,
pre-symptomatics,  and asymptomatics~\cite{Asymptomatic},
as well as hospitalizations, recoveries and fatalities. In addition,
we consider such populations in {\it both} the
unvaccinated and vaccinated portions of the population
and their interactions.
The primary aim of the associated study is to explore the
dynamical evolution of the omicron variant from
the end of 2021 to early spring 2022.
We also present a simpler model (model B) for the evolution of
the earlier delta variant during the fall
of 2021. In that case, vaccination was deemed to protect
individuals from being infected, and the fraction of
breakthrough infections was quite small,
even in groups such as
the potentially highly exposed group of
healthcare workers~\cite{10.1093/cid/ciab916}. Accordingly, we assume
that the vaccinated population may be
effectively removed from the susceptible compartment.

We examine two versions of the proposed omicron model in
Section~\ref{sec:Omicron} (models A1 and A2),
their difference motivated by data availability.  The population flows in
Model A1 terminate at the fatalities compartment: as such, the
model considers that fatalities in both the
unvaccinated and vaccinated populations provide the most reliable
data.
Model A2 is motivated by the existence of systematic data for the total number of
hospitalizations (conventional and ICU): here,
population flows terminate
at the hospitalizations compartment, i.e., they do not branch further
to the fatalities compartment as in model A1.
This is for a number of reasons: \emph{in primis}, reporting of fatalities occasionally occurs
retroactively (and less reliably).
Our models are applied
to two regions with similar populations (approximately
8 million inhabitants): Andalusia and Switzerland (including the Principality
of Liechtenstein)
The motivation for this choice arises, once again, from the availability
of suitably stratified data, whereby both fatalities and hospitalizations
are available for vaccinated and unvaccinated individuals.
In Section~\ref{sec:Delta} we propose,  in addition to
the more detailed model for the omicron variant presented in Section~\ref{sec:Omicron},
a model for the delta variant, model B.  We use model B in the same
spatial regions. We typically find that
numerical results compare favorably to available
data, both in terms of the comparison of the regression
results and also in connection to testing beyond the
end of the training period for the model parameters.
Finally, in Section~\ref{sec:Conclusions}, we summarize our findings and
present our conclusions.
In the Appendix, we
consider mathematically the question of structural identifiability
of the models developed herein.

\section{Omicron variant}
\label{sec:Omicron}
\subsection{Model A1: Branches terminate at fatalities}
\label{sec:ModelA1}

The first model for the omicron variant, model A1, extends our
compartmental epidemiological model
used to examine the COVID-19 pandemic evolution in Mexico~\cite{cuevas2021}. Accordingly, the susceptible
population $S$ can turn to exposed ($E$) through interactions
with either symptomatically infected ($I$), presymptomatic
($P$), or asymptomatic ($A$) individuals. The exposed population
$E$, in turn,
can convert to either $P$,  within a time scale $1/\sigma_1$,
leading to different clinical
stages of the disease or to $A$, a compartment that has been recognized
to play a key role in the dynamical evolution of COVID-19~\cite{Asymptomatic}. The asymptomatic
population $A$ can only lead to undisclosed recoveries (denoted as $U$), over a time scale $1/\mu$.
On the other hand, the presymptomatic individuals $P$
turn to infected with clinical symptoms $I$ over a time scale $1/\sigma_2$. The addition
of the latency period $1/\sigma_1$ and the preclinical
period $1/\sigma_2$ constitute the incubation time scale
of the disease, $\tau_{\textrm{inc}} = 1/\sigma_1 + 1/\sigma_2$.
Subsequently, the symptomatically infected can either turn to hospitalized $H$ at
a rate $\gamma_h$, while the rest may recover ($R$) at a rate $\gamma_r$.
Finally, those in the hospitalized population of $H$ can, again, branch into
two populations:
they either recover at
a rate $\kappa_r$, or they lead to fatalities ($D$)
at a rate $\kappa_d$.
\begin{figure}[htb]
    \centering
    \includegraphics[width=0.45\textwidth]{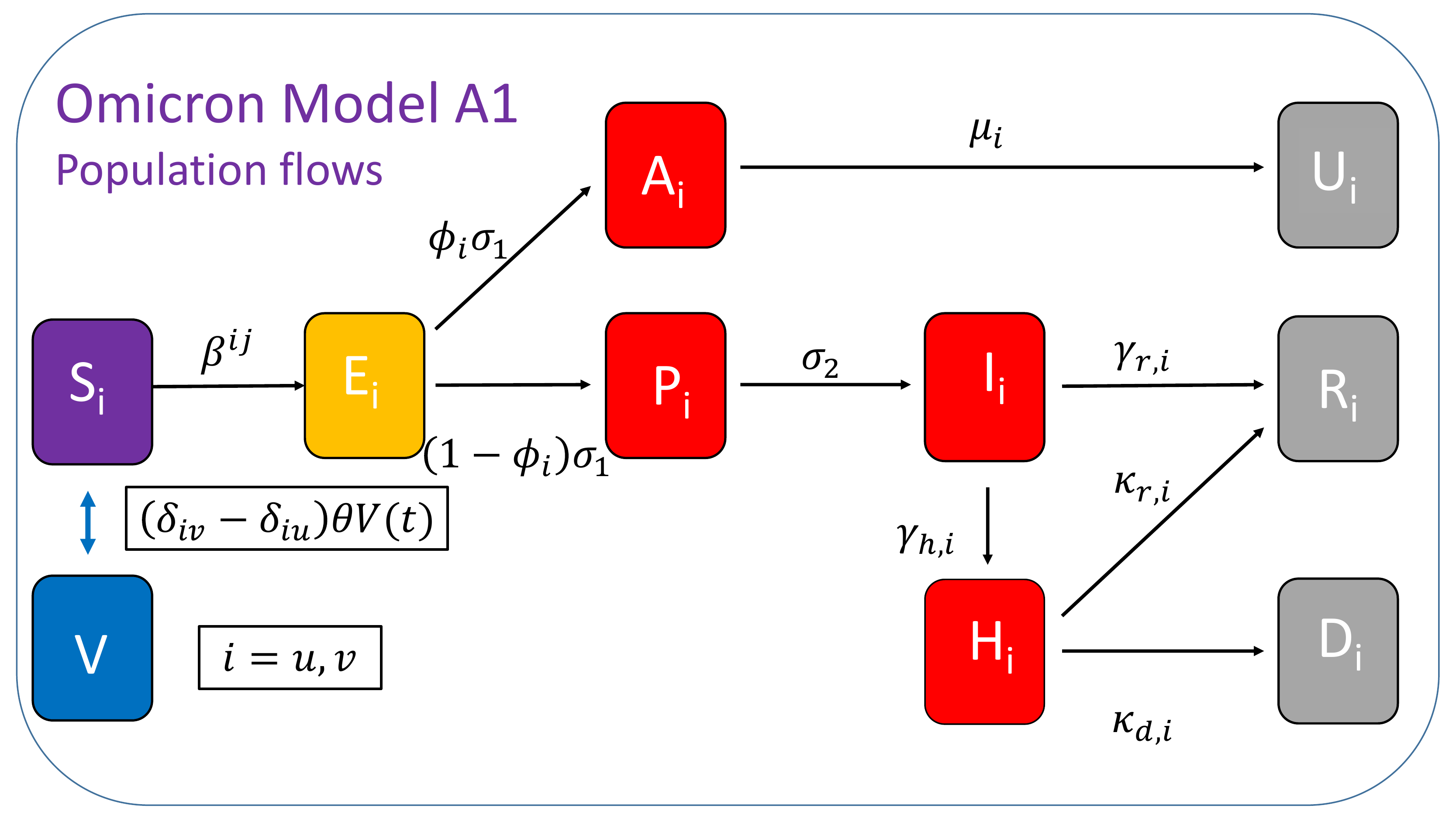}
    \hspace{1.5cm}
    \includegraphics[width=0.45\textwidth]{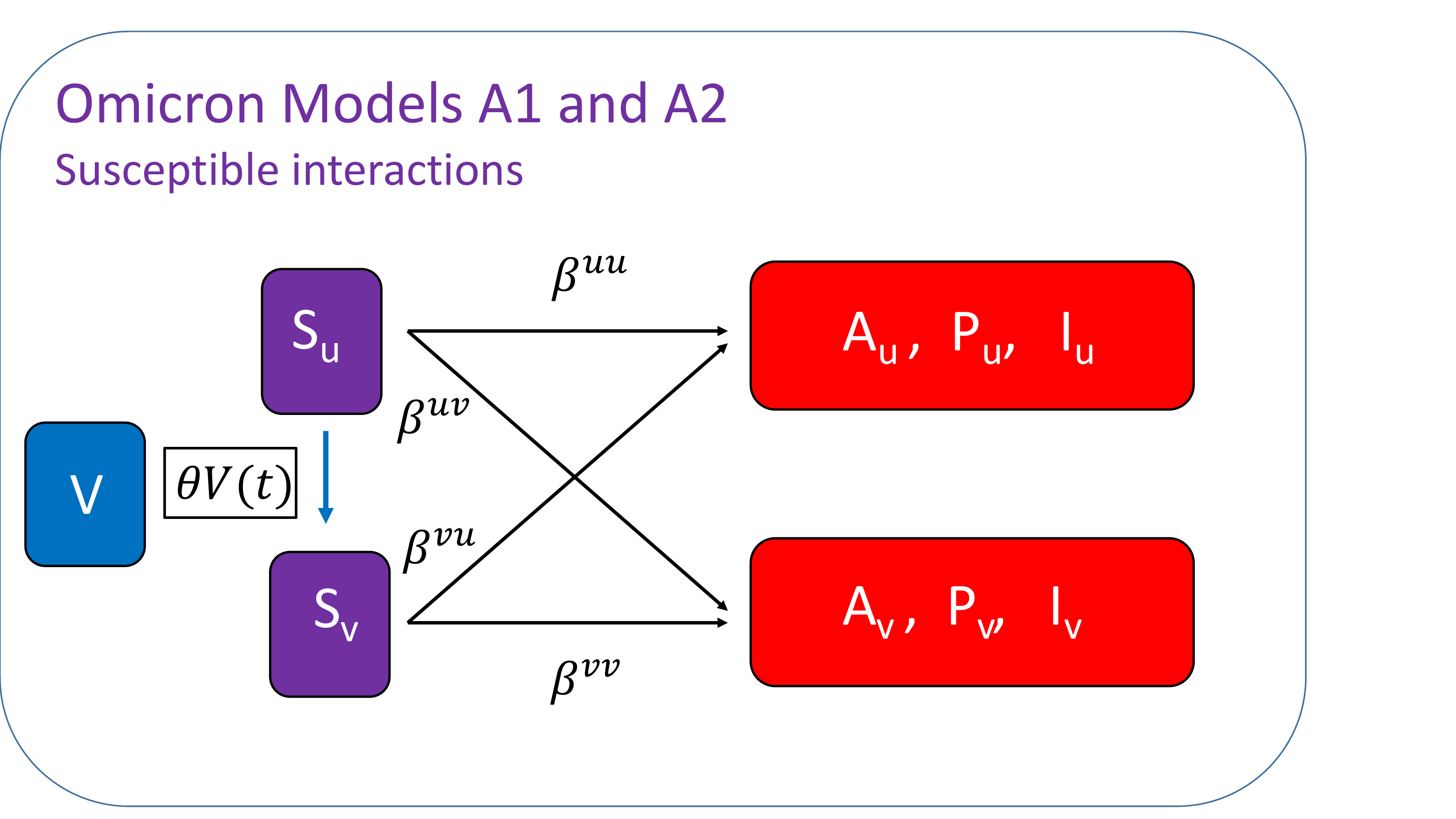}
    \caption{Schematic diagram of population flows according
    to model A1 (left panel) and susceptible interactions with other population compartments, for both models A1 and A2 (right panel).
    The symbol $\delta_{ij}$ with $i,j = u,v$ is the Kronecker delta.}
    \label{fig:ModelA1}
\end{figure}

While these populations were also present in our earlier
work~\cite{cuevas2021}, it is relevant to highlight
the differences of the omicron-variant modeling.
For the period under consideration (fall 2021 to spring 2022),
vaccines had been deployed extensively in
Andalusia and Switzerland.  More importantly,
breakthrough infections due to the omicron variant were substantial
within the vaccinated population (contrary
to the case of the delta variant considered in Section~\ref{sec:Delta}).
In light of that, we formulated {\it two} sets of
populations: one representing unvaccinated individuals, denoted
by (the subscript) $u$,
and the other representing the substantial population of
vaccinated individuals, denoted by (the subscript) $v$.
Each population
subset had its own set of parameters.
This division of the whole population into two
almost independent subgroups (they interact only through the vaccinated
time series) is reflected in the
presentation of the model schematic in Fig.~\ref{fig:ModelA1}.
In fact, Fig.~\ref{fig:ModelA1} summarizes population
flows and interactions: the left panel illustrates the main population compartments and
the corresponding flows, while the right
panel shows the interactions of susceptibles with other compartments, both
for vaccinated and unvaccinated populations. The model equations are:
\begin{subequations}
\begin{align}
\frac{\mathrm{d}S_u}{\mathrm{d}t} &= -\beta^{uu}S_u(I_u+A_u+P_u)-\beta^{uv}S_u(I_v+A_v+P_v)-\theta V(t)  , \label{Su1} \\
\frac{\mathrm{d}E_u}{\mathrm{d}t} &= -\sigma_1E_u+\beta^{uu}S_u(I_u+A_u+P_u)+\beta^{uv}S_u(I_v+A_v+P_v) , \label{Eu1} \\
\frac{\mathrm{d}P_u}{\mathrm{d}t} &= (1-\phi_u)\sigma_1E_u-\sigma_2P_u , \label{Pu1} \\
\frac{\mathrm{d}A_u}{\mathrm{d}t} &= \phi_u\sigma_1E_u-\mu_uA_u , \label{Au1} \\
\frac{\mathrm{d}U_u}{\mathrm{d}t} &= \mu_uA_u , \label{Uu1} \\
\frac{\mathrm{d}I_u}{\mathrm{d}t} &= \sigma_2P_u-(\gamma_{r,u}+\gamma_{h,u})I_u ,  \label{Iu1} \\
\frac{\mathrm{d}H_u}{\mathrm{d}t} &= \gamma_{h,u}I_u-(\kappa_{r,u}+\kappa_{d,u})H_u ,  \label{Hu1} \\
\frac{\mathrm{d}R_u}{\mathrm{d}t} &= \gamma_{r,u}I_u+\kappa_{r,u}H_u , \label{Ru1} \\
\frac{\mathrm{d}D_u}{\mathrm{d}t} &= \kappa_{d,u}H_u ,  \label{Du1} \\
\frac{\mathrm{d}S_v}{\mathrm{d}t} &= -\beta^{vv}S_v(I_v+A_v+P_v)-\beta^{vu}S_v(I_u+A_u+P_u)+\theta V(t) ,  \label{Sv1}  \\
\frac{\mathrm{d}E_v}{\mathrm{d}t} &= -\sigma_1E_v+\beta^{vv}S_v(I_v+A_v+P_v)+\beta^{vu}S_v(I_u+A_u+P_u) , \label{Ev1}\\
\frac{\mathrm{d}P_v}{\mathrm{d}t} &= (1-\phi_v)\sigma_1E_v-\sigma_2P_v ,  \label{Pv1} \\
\frac{\mathrm{d}A_v}{\mathrm{d}t} &= \phi_v\sigma_1E_v-\mu_vA_v , \label{Av1} \\
\frac{\mathrm{d}U_v}{\mathrm{d}t} &= \mu_vA_v , \label{Uv1} \\
\frac{\mathrm{d}I_v}{\mathrm{d}t} &= \sigma_2P_v-(\gamma_{r,v}+\gamma_{h,v})I_v ,  \label{Iv1} \\
\frac{\mathrm{d}H_v}{\mathrm{d}t} &= \gamma_{r,v}I_v-(\kappa_{r,v}+\kappa_{d,v})H_v ,  \label{Hv1} \\
\frac{\mathrm{d}R_v}{\mathrm{d}t} &= \gamma_{h,v}I_v+\kappa_{r,v}H_v ,  \label{Rv1} \\
\frac{\mathrm{d}D_v}{\mathrm{d}t} &= \kappa_{d,v}H_v  \label{Dv1}.
\end{align}
\label{ceq1}
\end{subequations}

We made a
number of simplifying assumptions to reduce
the number of parameters and enhance
the identifiability of the model (see also the relevant analysis
in Appendix~\ref{sec:IA}).
We consider a model with only four transmission rates $\beta^{ij} (i,j = u,v)$.
We assumed that infectious contacts could only occur
between four groups: between unvaccinated individuals (unvaccinated-unvaccinated
contacts denoted by the superscript $uu$), between vaccinated and vaccinated individuals
(denoted by the superscript $vv$) and across these
two groups (denoted by $uv$ ---for vaccinated transmitting
to unvaccinated and $vu$ for the reverse path of infection).
Notice that $uv$ and $vu$ are not \emph{a priori} assumed to be equivalent.
Within each subgroup of infection transmission
($uu, vv$,  $uv$, and $vu$), the infections induced by the three infectious compartments
$P$, $A$ and $I$ are assumed to occur at the same rate, i.e., the transmission rate
is considered to be independent of whether the infectious individual
exhibits symptoms ($I$) or not ($A,P$).
While we expect these transmission rates to differ (in fact, we know that even within
a given population the $S - I$ transmission rate differs from the $S - A$
transmission rate, see for example, Ref.~\cite{cuevas2021}), the
identifiability analysis based on the available time series suggests that they would not
be independently computable in a definitive way.
In addition, we introduced a single constraint that requires that the incubation period of the disease
$\tau_{\textrm{inc}} = \sigma_1^{-1}+\sigma_2^{-1}$ be a value randomly sampled from a
normal distribution with mean 3.42 and standard deviation 0.2755. This
leverages information about the (shorter) incubation period associated
with the omicron variant~\cite{jama2022}.

Vaccine efficiency is introduced via the parameter $\theta$,
whose variation bounds were set in the range 75\%--95\%. This factor multiplied
by the time series of vaccinations $V(t)$ effectively ``transfers'' individuals from the unvaccinated susceptible
population to the vaccinated susceptible compartment.
It is important to remark that $H$ measures both conventional and critical hospitalizations together.
While we recognize the relevance of the ongoing debate of distinguishing
 deaths ``from COVID'' vs.
``with COVID''~\cite{Slatere189}, unfortunately the data available
herein do not allow for a definitive distinction between the two.

We obtained the best-fit parameters and initial conditions
by minimizing an appropriately chosen
norm.  For both regions of interest,
the time period used for the fits was from November 15, 2021 to March 1, 2022.
The identification of the date a particular variant appeared in a geographical
location is fraught with uncertainties.The choice of November 15, 2021 as the initial day of fittings stems
from a number of indirect indications: Ref.~\cite{Germany_Omicron}  reports a surge of cases in Germany at the beginning of November;
Ref.~\cite{TheNetherlands_Omicron} mentions that an omicron-variant case was reported on November 19; and
the WHO site~\cite{WHO_Omicron} mentions that in South Africa the first confirmed infection was
reported on November 24,
although arising in
the sequencing of a sample collected on November 9.
Additionally, inspection of the data shows a gradual increase starting at November 15, after a plateau.
The effective parameter training period indicated
above (till March 1, 2022),  is followed by a prediction
period (with the optimal parameters and initial conditions fixed, as determined in the training period).
The predicted time series that terminates on March 29, 2022 is then compared to the reported data.
Predictions do not go beyond that date because the measurement strategy in Andalusia changed,
thereby rendering our fixed parameters of limited relevance to the new data.
Moreover, around that time Spanish public policy also changed, and face masks
were no longer required.  In Switzerland some restrictions were removed in
the middle of February. More details are presented in the appropriate
results sections.

We perform two separate fits, i.e., we used two different norms to compare
predictions to reported numbers depending on data availability.
First, we fit the predicted total number of fatalities to the reported number
by minimizing the norm $\mathcal N$ (i.e., the loss function)
\begin{equation}\label{eq:N_DTotal}
    {\mathcal N}=\frac{1}{n}\sum_{i=1}^n\Big\{\log\big[D_{u,\mathrm{num}}(t_i)+D_{v,\mathrm{num}}(t_i)\big]-\log\big[D_{\mathrm{obs}}(t_i)\big]\Big\}^2 ,
\end{equation}
where the subscript ``$\textrm{num}$" refers to predicted (calculated) numbers
and ``$\textrm{obs}$" to observation (reported numbers), and $n$ refers to the number (days) of observations.
This loss function effectively does not \emph{distinguish} the compartmental
origin of the fatalities, i.e., whether they arise from the $u$ or $v$
compartments: it only accounts for the  cumulative number of fatalities.
This will, inevitably, result in the determination of some parameters
between the unvaccinated and the vaccinated populations
that may not necessarily be epidemiologically meaningful (as we
will see in the detailed comparisons of our predictions for Andalusia
and Switzerland).

Whenever we used norm~(\ref{eq:N_DTotal}), we also
included the waning effect of the vaccines and booster vaccination
effects, in addition to those fully vaccinated.
It is well-documented that
different vaccines have different waning immunities
(see, for instance, the detailed analysis of Ref.~\cite{healthdata}).
However, to be able to account for these
effects without adding a large number of additional coefficients,
we assumed that vaccines are roughly effective for
an interval of about 180 days.\ Consequently, we define $V(t)$ as:
\begin{equation} \label{eq:vaccineA1}
    V(t)=V_{fv}([t])-V_{fv}([t-180])+V_b([t]) ,
\end{equation}
where the subscripts $fv$ refers to fully vaccinated, and $b$ to booster.

The above optimization via norm~(\ref{eq:N_DTotal}) is a point estimator,
that is, a single set of parameters and initial conditions is obtained.
To calculate their confidence intervals, and consequently
the confidence interval of the predictions, we follow
the bootstrapping method described in \cite{Chowell}.  The first step is to generate 250 random, synthetic,  time series for
the fatalities based on the reported data.
To accomplish this, we first apply the optimization to find the best fit to
the original data set: we refer to that optimization of the reported fatalities
data as the ``numerical truth".
In this first optimization, we also included $I_u(0)$, $I_v(0)$, $E_u(0)$, $E_v(0)$, $A_u(0)$, $A_v(0)$, $H_u(0)$ and $H_v(0)$
as fitting parameters: these parameters were fixed in the subsequent bootstrapping steps.
Then, random noise of a prescribed level,
empirically chosen to be 5\%,
was added to the ``numerical truth"
to generate 250 ``polluted'' (i.e., noisy) time series for the fatalities.
Knowledge of the error in data collection may be helpful to select an appropriate
noise level.
In the second step, we apply the optimization procedure to find the best fit to each of the 250
synthetic fatalities time series to obtain 250 sets of parameters
from which the confidence intervals for the parameters and predictions can be computed.
The same bootstraping procedure was used for the hospitalization time series.

We also fitted separately, if the reported data allowed us,
the vaccinated and unvaccinated
fatalities time series using them as separate inputs to
our minimization objective. In that case, the relevant norm is
\begin{equation}\label{eq:N_DuDv}
     {\mathcal N}=\frac{1}{n}\sum_{i=1}^n \Big \{ \log \big [ D_{u,\mathrm{num}}(t_i)\big ] -
     \log \big [ D_{u,\mathrm{obs}}(t_i) \big ] \Big \}^2
    +\Big \{ \log \big [ D_{v,\mathrm{num}}(t_i) \big ]
    - \log \big [ D_{v,\mathrm{obs}}(t_i) \big ]  \Big \}^2 .
\end{equation}
With this norm, we are genuinely treating the vaccinated compartment
separately: we expect its fraction of
fatalities (proportionally to the
corresponding susceptible population)
to be reflected in the obtained parameters.
It should be added that in this case, given the way that the
data are obtained, the vaccinated status
 corresponds to people who had received the full doses, independently of
 antibodies waning, boosting, or efficacy of vaccines. Consequently, $\theta$ has been fixed to 1 in every fit, and $V(t)$ is defined as
\begin{equation} \label{eq:vaccineFull}
    V(t)=V_{fv}([t]) .
\end{equation}

We mention here that an important consideration pertinent to the model
concerns the identifiability of its coefficients (and initial conditions).
This pertains to whether, based on the time series given, the unknown
model parameters can be uniquely identified~\cite{Eisenberg2013}. In addition to the question whether all parameters can be uniquely identified (global identifiability)
or some may have multiple possible values  (local identifiability),
there are also practical
issues concerning whether different sets of parameters lead to similar
(although not necessarily identical) observations; see, e.g., the
discussion of~\cite{Identify_PINN}. Here,
following the approach presented in Appendix~\ref{sec:IA}
(see also~\cite{Pogudin_SIAN,Pogudin_2021,Pogudin_2022}), we find that all the parameters are globally identifiable,
except
$$
\gamma_{h, u},  \quad \gamma_{r, u}, \quad \gamma_{h, v} ,  \quad \gamma_{r, v}, \quad \kappa_{d,u}, \quad \kappa_{r,u},
\quad \kappa_{d, v}, \quad \kappa_{r,v} .
$$
 For these eight parameters, the following combinations are globally identifiable:
$$
\gamma_{h, u} + \gamma_{r, u}, \quad \gamma_{h, v} + \gamma_{r, v},  \quad \kappa_{d, u} + \kappa_{r, u}, \quad \kappa_{d, v} + \kappa_{r, v},
 \quad \gamma_{h, u} \kappa_{d, u},  \quad \gamma_{h, v} \kappa_{d, v}.
$$
As concerns the initial conditions,  $H_u(0)$ and $ H_v(0)$ are not identifiable,  in addition to the initial conditions
for the terminal compartments $U_u, U_v, R_u, R_v$.

  \begin{figure}[htb]
    \centering
    \includegraphics[width=0.5\textwidth]{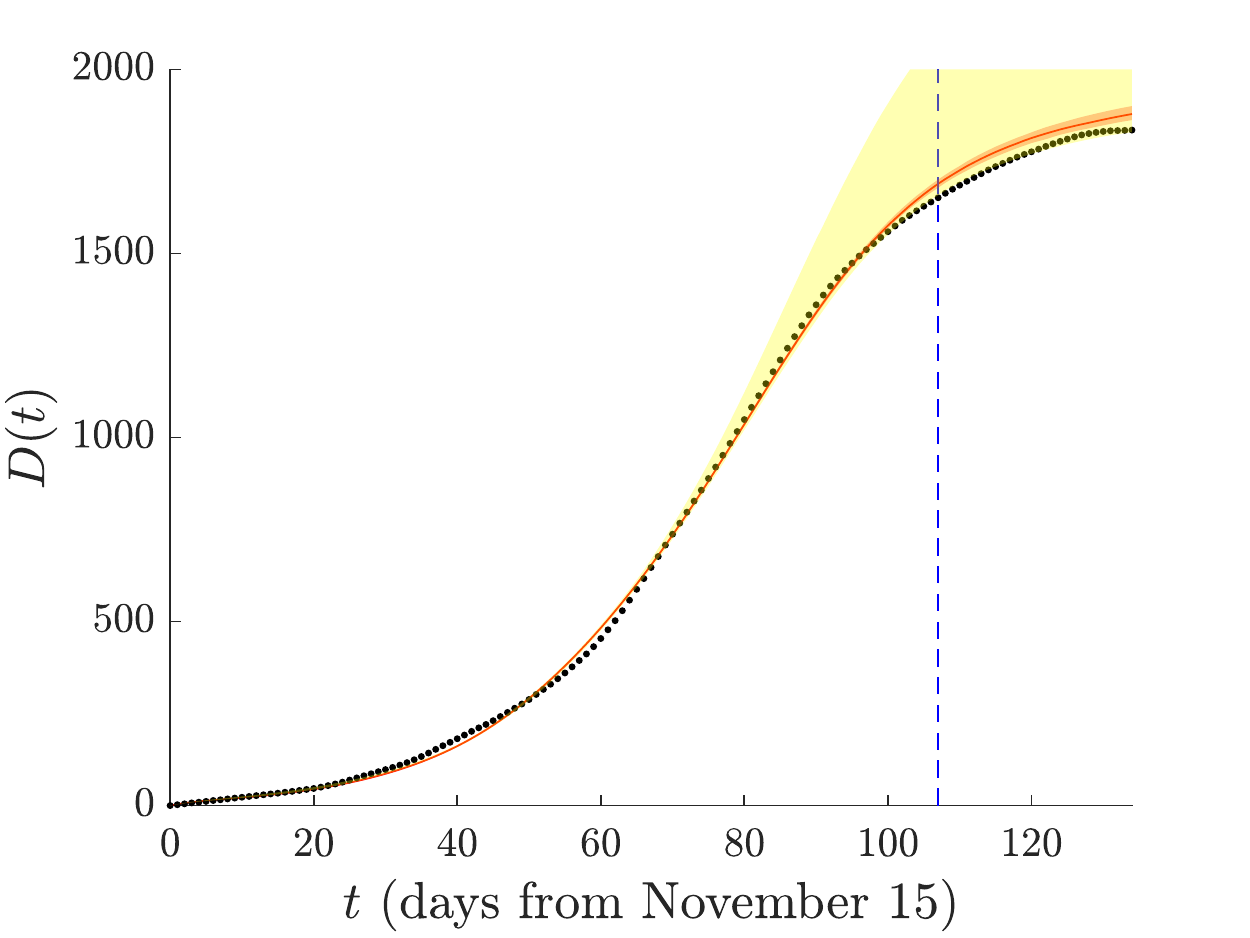}
    \caption{Omicron-variant model A1: Fit and prediction for the total number
    of fatalities in Andalusia (norm  of Eq.~(\ref{eq:N_DTotal})). The calculated curve is plotted in red along
    with its confidence/prediction intervals: red shade corresponds to the
    interquartile range,  while the yellow shade corresponds to the 95\% confidence interval comprised between the 2.5 and 97.5 percentiles.
    Reported data for the total number of fatalities are given by the black points.
    The vertical line, the beginning of the prediction interval, is March 1, 2022.}
    \label{fig:total_deaths}
\end{figure}

The identifiability analysis suggests that $\phi_u, \phi_v$ and many other parameters are globally identifiable, i.e., they have a unique value
given the functions $D_u(t)$ and $D_v(t)$.
However, these theoretical-analysis results do not exactly transfer to numerical calculations for various reasons.
A globally identifiable parameter may not necessarily have a sharp estimate due to the potential sloppiness~\cite{Sloppy_2007} of the model.
When the output functions are not sensitive to a particular parameter, a sharp estimate will not be expected, even though the parameter may be
globally identifiable. Many other factors, e.g., the reliability and accuracy of the reported time series,  may  exacerbate the situation.
The identifiability analysis assumes that both functions $D_u(t)$ and $D_v(t)$ are outputs, which includes much more information than the simple
loss term Eq.~\eqref{eq:N_DuDv} can provide.
It is an open question how close these estimates are to the actual parameters.
Interestingly, however, we find that
these estimates still give fairly accurate predictions, even though certain parameter estimates may not be as sharp as desired.
When the total death, $D_u(t)+D_v(t)$, is the only output, we cannot obtain any identifiability results.
The implications of there remarks on model identifiability are further elaborated in
our comments of best-fit parameters.

\subsubsection{Andalusia}
\label{sec:AndalusiaA1}

\begin{table}[htb]
\begin{center}
\caption{\label{tab:andalusia}
Optimal parameters and initial conditions for the
omicron-variant models in Andalusia: Model fits to the total number of fatalities,
discussed in Section~\ref{sec:AndalusiaA1} (model A1, third column)
and to the total number of hospitalizations, discussed in Section~\ref{sec:AndalusiaA2} (model A2, right column).
{Population $N_{\textrm{And}} = 8.4M$.}}
\begin{tabular}{c|c|c|c}
\hline \hline
Parameter & Symbol & Median (interquartile range) & Median (interquartile range) \\
\hline
& & Fit to total number of deaths & Fit to total number of hospitalizations \\
& & [Model A1, norm Eq.~(\ref{eq:N_DTotal})] &
[Model A2, norm Eq.~(\ref{eq:N_JTotal})] \\ \hline
Transmission rate $uu$ [per day] & $\beta^{uu}$ & 0.6437 (0.6313--0.6532) & 0.1279 (0.1134--0.1491)\\
Transmission rate $uv$ [per day] & $\beta^{uv}$ & 0.3687 (0.3040--0.4044) & 0.0209 (0.0166--0.0270)\\
Transmission rate $vu$ [per day] & $\beta^{vu}$ & 0.0679 (0.0591--0.0822) & 0.2400 (0.1775--0.2855)\\
Transmission rate $vv$ [per day] & $\beta^{vv}$ & 0.1939 (0.1839--0.2168) & 0.4516 (0.4381--0.4637)\\
Latency period [days] & $1/\sigma_1$ & 1.7923 (1.6999--1.8935) & 1.8143 (1.7266--1.9164)\\
Preclinical period [days] & $1/\sigma_2$ & 1.6467 (1.5513--1.7496) & 1.6300 (1.5038--1.7509)\\
$A_u/P_u$ partitioning & $\phi_u$ & 0.3689 (0.3639--0.3832) & 0.3592 (0.3486--0.3743)\\
$A_u/P_u$ partitioning & $\phi_v$ & 0.3363 (0.3311--0.3457) & 0.3933 (0.3816--0.4063)\\
Infectivity period ($A_u$) [days] & $1/\mu_u$ & 2.9185 (2.8497--2.9433) & 3.1540 (3.0764--3.2333)\\
Recovery rate $I_u \rightarrow R_u$ [[per day] & $\gamma_{r,u}$ & 0.1999 (0.1958--0.2103) & 0.1998 (0.1871--0.2082)\\
Transition rate $I_u \rightarrow H_u$ [per day] & $\gamma_{h,u}$ & 0.0056 (0.0050--0.0060) & 0.0068 (0.0062--0.0072)\\
Infectivity period ($A_v$) [days] & $1/\mu_v$ & 3.2038 (3.1581--3.2467) & 3.3018 (3.1663--3.6227)\\
Recovery rate $I_v \rightarrow R_v$ [per day] & $\gamma_{r,v}$ & 0.1728 (0.1692--0.1802) & 0.2088 (0.2023--0.2153)\\
Transition rate $I_v \rightarrow H_v$ [per day] & $\gamma_{h,v}$ & 0.0057 (0.0051--0.0064) & 0.0015 (0.0014--0.0017)\\
Recovery rate $H_u \rightarrow R_u$ [per day] & $\kappa_{r,u}$ & 0.3436 (0.3314--0.3576) & --- \\
Death rate $H_u \rightarrow D_u$ [per day] & $\kappa_{d,u}$ & 0.0090 (0.0080--0.0106) & --- \\
Recovery rate $H_v \rightarrow R_v$ [per day] & $\kappa_{r,v}$ & 0.3354 (0.3198--0.3506) & --- \\
Death rate $H_v \rightarrow D_v$ [per day] & $\kappa_{d,v}$ & 0.0097 (0.0092--0.0101) & --- \\
Vaccine efficieny [-] & $\theta$ & 0.8825 (0.8767--0.8885) & 0.8510 (0.8436--0.8598)\\
\hline
Initial condition & & & \\
\hline
Initial unvaccinated Exposed ($E_u$) population [\#] & $E_u(0)$ & 641 & 1485 \\
Initial unvaccinated Presymptomatic ($P_u$) population [\#] & $P_u(0)$ & 2217 & 1541 \\
Initial unvaccinated Asymptomatic ($A_u$) population [\#] & $A_u(0)$ & 2273 & 1124 \\
Initial unvaccinated symptomatically Infected ($I_u$) population [\#] & $I_u(0)$ & 2677 & 2438 \\
Initial unvaccinated Hospitalized ($H_u$) population [\#] & $H_u(0)$ & 60 & ---  \\
Initial vaccinated Exposed ($E_v$) population [\#]  & $E_v(0)$ & 840 & 3218 \\
Initial vaccinated Presymptomatic ($P_v$) population [\#] & $P_v(0)$ & 2902 & 3339 \\
Initial vaccinated Asymptomatic ($A_v$) population [\#] & $A_v(0)$ & 2976 & 2435 \\
Initial vaccinated symptomatically Infected ($I_v$) population [\#] & $I_v(0)$ & 3504 & 5281 \\
Initial vaccinated Hospitalized ($H_v$) population [\#] & $H_v(0)$ & 370 & ---  \\
\hline \hline
\end{tabular}
\end{center}
\end{table}
As mentioned, the fitting time window we used to obtain
the optimized parameters and initial conditions was
from November 15, 2021  to March 1, 2022. The prediction interval
ended on March 29, 2022.
The time series for Andalusia is available from the  Spanish Health Ministry,
but we used the series compiled at \cite{TimeSeriesAndalusia}.
Note that in Andalusia the reported values of $D(t)$, and the total number of
hospitalizations $J(t)$, the latter discussed in Section~\ref{sec:ModelA2}, correspond to the event day,
whereas for Switzerland they correspond to the report day.

The vaccination data until April 29, 2021 were also taken from \cite{TimeSeriesAndalusia}.
After that date, they
were extracted from the Regional Government of Andalusia
(Junta de Andaluc\'{\i}a,  Ref.~\cite{AndalusiaVaccine}).
We ignored the vaccinations for kids under 12 years old, as there were many
data anomalies, resulting in a time series that appears to be
problematic.
Irrespective of that, this population segment corresponds to only $\lesssim4$\% of the total vaccinations.
The fatalities time series we used did not report
how many fatalities could be attributed to vaccinated or unvaccinated individuals.
Therefore, for the region of Andalusia we used only norm~(\ref{eq:N_DTotal}),
coupled to the modified vaccination time series as described in Eq.~(\ref{eq:vaccineA1}), to perform the optimizations.

Figure \ref{fig:total_deaths} shows the calculated
fatalities time series (both fitting and prediction intervals) and
the reported numbers. Table \ref{tab:andalusia} (model A1 in column 3)
presents the optimized parameters and initial conditions.
The  reported interquantile range arises from 250 fits in the bootstraping step, as discussed above.

We observe that the overall trend of the fatalities seems
to be reasonably well captured by the model within its prediction
intervals (and their associated uncertainty). We do note, however, a
slight over-prediction towards the end of the time series,  during March 2022.
The transmission rates $\beta^{ij}$ ($i,j = u,v$)
with at least one
member of the unvaccinated population $uu$ and $uv$  are clearly higher
than the $vv$ rate between
members of the vaccinated population. In fact,
$\beta^{uu}$ is more that three times higher
than $\beta^{vv}$.  The lowest transmission rate
is predicted to be $\beta^{vu}$, even lower that $\beta^{vv}$.
At this point it is important to recall our discussion about the use of the total
number of deaths and the resulting inability to identify definitively
the model parameters. Hence, the above numbers, even
when they appear to be intuitively relevant, should be taken
with a grain of salt.
The latency period is approximately 3.5 days,  as imposed by our
constraint, and in agreement with~\cite{jama2022}.
The role of asymptomatics, as reflected by the fraction $\phi_i$
of exposed who become asymptomatics,
is considerable, approximately 1/3 and independent of whether
the population is vaccinated or not.  The calculated fraction of
asymptomatics is in reasonable agreement with Ref.~\cite{AsymptomaticsReview2022}
who reported a pooled fraction of asymptomatics for the omicron
variant of 25.5\% (95\% confidence interval 17.0\% -38.2\%).
The vaccinated and unvaccinated infectivity period {for asymptomatic infections} $1/\mu_i$ is
approximately constant, at about three days, again independent of
whether the $u$ or $v$ compartment is considered.

We also find that some parameters are more difficult to justify,
Specifically, we find that the recovery rates of symptomatically
infected individuals $I_i \rightarrow R_i$ and that of the hospitalized
individuals $H_i \rightarrow R_i$ are
almost independent of whether the population is vaccinated or not.
The same holds for the transition rates $I_i \rightarrow H_i$
and the death rates $H_i \rightarrow D_i$:
all four of them are found to have weak variations.
The independence
of these rates on the administration of the vaccine might
be related to the norm we used that does not distinguish
between fatalities of vaccinated or unvaccinated individuals. We will return
to this point in our analysis of the Switzerland data.

\subsubsection{Switzerland}
\label{sec:SwitzerlandA1}

\begin{figure}[htb]
    \centering
    \includegraphics[width=0.5\textwidth]{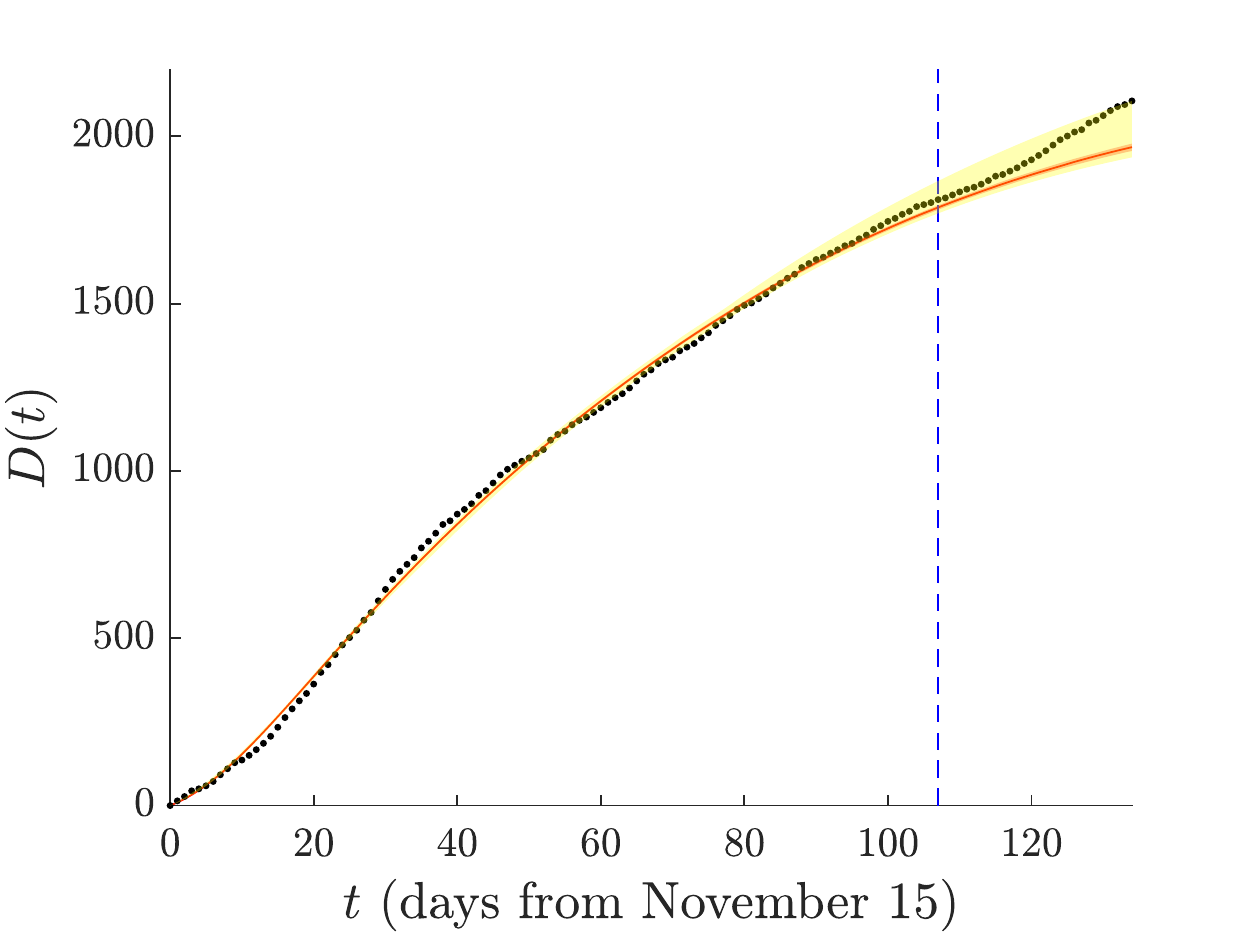}
     \caption{Omicron-variant model A1: Fit and prediction for the total number
    of fatalities in Switzerland (norm  Eq.~(\ref{eq:N_DTotal})). The calculated curve is plotted in red along
    with its confidence/prediction intervals: red shade corresponds to the
    interquartile range,  while the yellow shade corresponds to the 95\% confidence interval comprised between the 2.5 and 97.5 percentiles.
    Reported data for the total number of fatalities are given by the black points.
    The vertical line, the beginning of the prediction interval, is March 1, 2022.}
    \label{fig:total_deaths_chfl}
\end{figure}

\begin{figure}[htb]
    \begin{center}
    \begin{tabular}{cc}
    \includegraphics[width=0.5\textwidth]{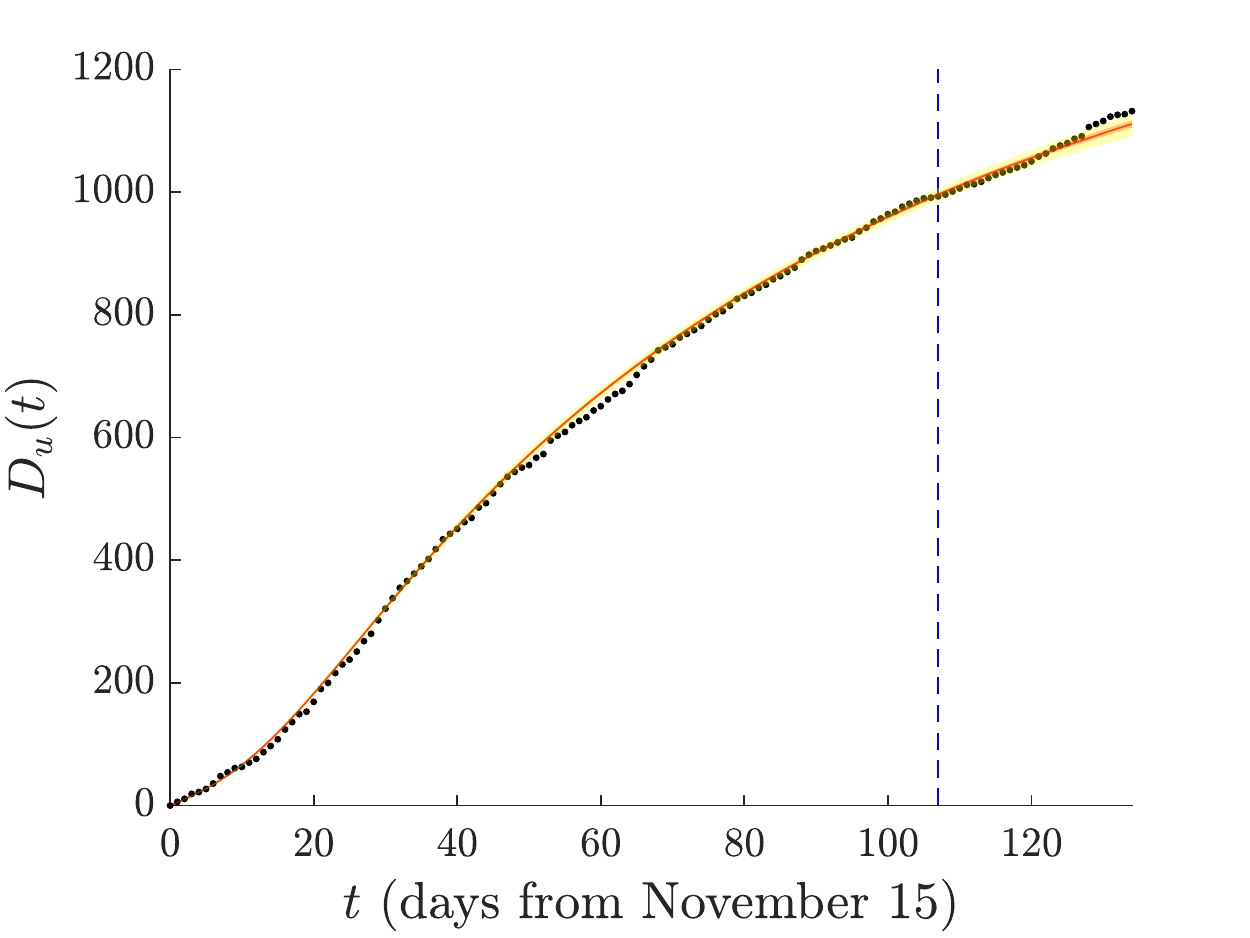} &
    \includegraphics[width=0.5\textwidth]{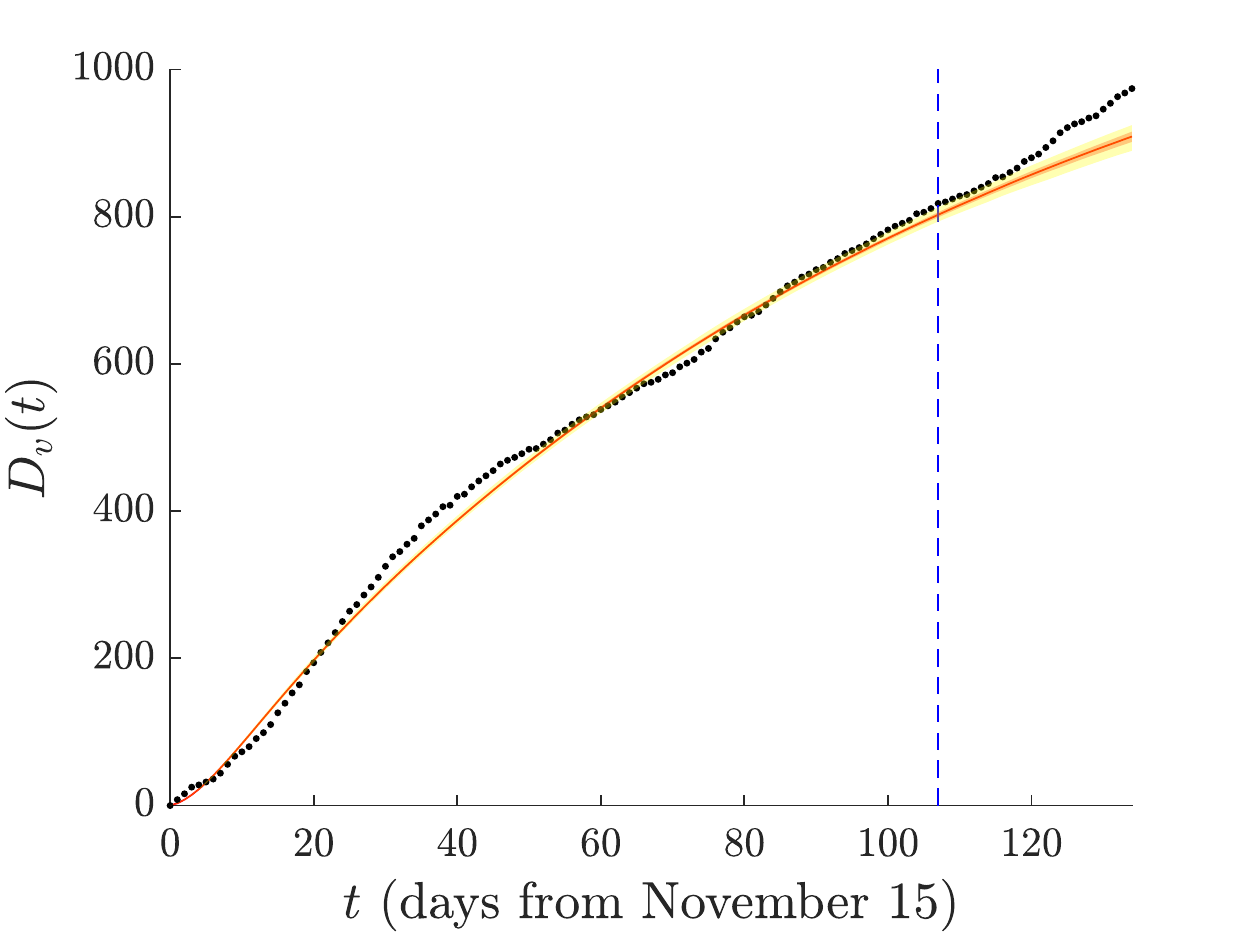} \\
    \end{tabular}
    \end{center}
        \caption{Omicron-variant model A1: Separate fits of vaccinated and
        unvaccinated fatalities in Switzerland
    (norm  Eq.~(\ref{eq:N_DuDv})).  Left panel: Fatalities of vaccinated individuals.
    Right panel: Fatalities of unvaccinated individuals.
    Calculated curves are plotted in red along
    with its confidence/prediction intervals: red shade corresponds to the
    interquartile range, whereas the yellow shade presents the 95\% confidence interval comprised between the 2.5 and 97.5 percentiles.
    Reported data for the total number of vaccinated and unvaccinated fatalities are given by the black points.
    The vertical line, the beginning of the prediction interval, is March 1, 2022.}
    \label{fig:deaths_chfl}
\end{figure}

We chose to perform model calculations for a
territory with a population similar in
number to that of Andalusia, and for which adequate data are available.
As such, we chose a region that contains Switzerland and the Principality of
Liechtenstein (data are jointly reported) since the total population of this
aggregate territory is 8.7M, (compared to 8.4M  for Andalusia).
Overall, we followed a procedure very similar to
what we used for Andalusia, with a few minor changes.
Identical fitting and prediction intervals are used as those for
Andalusia. We do note, however, that starting February 17, 2022 most
restrictions were lifted in Switzerland. We believe this is one of the reasons
we observe a model under-prediction of the number of fatalities
in Fig.~\ref{fig:total_deaths_chfl} and in Figs.~\ref{fig:deaths_chfl}.

Case reporting was slightly different.
The Swiss government through the Federal Office of Public Health
provides daily the status (vaccinated, unvaccinated or unknown) of each
hospitalized/deceased person at Ref.~\cite{OpenDataSwiss}.
In the absence of a concrete
metric on how to partition
unknown fatalities to
vaccinated and unvaccinated individuals, we used the following
procedure to convert
these three time series
into two,
one associated
with vaccinated $D_v(t_i)$ and the other to unvaccinated individuals $D_u(t_i)$.
Let $\bar{d}_v(t_i)$, $\bar{d}_u(t_i)$ and $\bar{d}_{\star}(t_i)$
denote the number of daily reported fatalities with vaccinated, unvaccinated, and unknown state, respectively.
We randomly sample an integer number $\delta_i\in[0,\bar{d}_{\star}(t_i)]$, following a uniform distribution, and then we define the daily number of vaccinated/unvaccinated deceased
as $d_v(t_i)=\bar{d}_v(t_i)+\delta_i$ and
$d_u(t_i)=\bar{d}_u(t_i)+ [\bar{d}_{\star}(t_i)-\delta_i]$.
The total number of deaths is the cumulative
sum, i.e. $D_v(t_i)=\sum_{j=1}^i d_v(t_j)$ and $D_u(t_i)=\sum_{j=1}^i d_u(t_j)$.
Note that we followed the same procedure to
generate the hospitalizations $J_v$ and $J_u$ used in model A2, in Section~\ref{sec:SwitzerlandA2}.
As mentioned earlier,  $D(t)$ and $J(t)$ for Switzerland correspond to the report day.

Given the reconstructed time series $D_i(t)$ we used norm~(\ref{eq:N_DuDv}), in
addition to the norm~(\ref{eq:N_DTotal}) used in the case of
Andalusia, to fit and predict
the fatalities time series for the territory of
Switzerland (and the Pricipality of Liechtenstein).  We attempted to fit separately
the vaccinated and unvaccinated
deceased, using them as separate inputs to
our minimization objective.
As mentioned earlier, since we consider
that vaccinated individuals have received the full dose (neglecting immunity
waning, boosting, of vaccine efficiency)
we take $\theta=1$ in every fit, and $V(t)$ is defined as
described in Eq.~(\ref{eq:vaccineFull}).
Our results for the fit to the total
number of fatalities are shown in Fig.~\ref{fig:total_deaths_chfl},
whereas those for the separate fits to vaccinated and unvaccinated deaths
are presented in Fig.~\ref{fig:deaths_chfl}.
Table~\ref{tab:switzerland}, columns two and four,
presents the fitting parameters and initial conditions.

We can see a  clear model under-prediction of the fatalities
(within the testing period),
for both optimizations (norm~(\ref{eq:N_DTotal}) and (\ref{eq:N_DuDv})),
despite an accurate following of the time-series trend throughout
the period over which regression is performed.  The
under-prediction is more severe in the case of the total number of
fatalities, Fig.~\ref{fig:total_deaths_chfl}, and in
the vaccinated death time series of the right panel in Fig.~\ref{fig:deaths_chfl}.
As mentioned earlier, we attribute the under-prediction
to the fact that after the end of the fitting period,
restrictions were considerably relaxed leading to more cases,
and eventually more fatalities, a feature that was not explicitly factored
in the model.

As regards the parameters of the model, we observe very similar
trends to what we obtained for Andalusia. A notable
exception is that in the total-deaths
fit $\beta^{uv}$ is the highest transmission rate,
retaining however $\beta^{uu} \gg \beta^{vv}$
in the case of norm~(\ref{eq:N_DTotal}). The latency period
is well reproduced (as expected due to the constraint and Ref.~\cite{jama2022}), and
the fraction of asymptomatics is approximately 25\% (again in agreement with~\cite{AsymptomaticsReview2022})
irrespective of vaccination or not. The remaining parameters
follow similar trends as reported in Table~\ref{tab:andalusia} for Andalusia.
It is noteworthy that in both cases recovery, transmission, and death rates
seem to depend relatively weakly on whether the vaccine had been administered or not.

A comparison of the parameters predicted by the two optimization is in order.
When the two distinct populations are used in the regression,
we observe, in the third column in
Table~\ref{tab:switzerland}, that the transition rate $\beta^{vv}$ becomes
the largest one with $\beta^{vu}$ being the smallest.
While a calculated higher viral transmissivity of vaccinated individuals
could, in principle, be attributed to
taking fewer measures to limit pathogen transmission via behavioral changes,
e.g., higher contact rates, negligence to use
face masks, etc, it is not obvious that such an attribution
is meaningful, rather than the potential outcome of the sloppiness
of the model.
Another surprising feature is that we do not find
a significant dependence of the parameters on the norm used
(apart from the noted difference in the transmission rates). The asymptomatic
fraction is predicted to be slightly larger, approximately 30\%, the
$H_u \rightarrow R_u$ is slightly smaller, and the death rate is slightly
larger.

\begin{table}[htb]
\begin{center}
\caption{\label{tab:switzerland}
Optimal parameters and initial conditions for the
omicron-variant models in Switzerland:
Model fits to (a) the total number of fatalities
 (model A1, second column from the left), (b) the total number of hospitalizations (model A2,  third column), (c) to the $u,v$ total number of fatalities separately
(model A1, fourth column), and (d) to the $u,v$ total number of hospitalizations
 separately (model A2,  fifth column).
 {Population $N_{\textrm{CHL}} = 8.7M$.  Parameters descriptions are as
 defined in Table~\ref{tab:andalusia}.}}
\begin{tabular}{c|c|c|c|c}
\hline \hline
Parameter & Median (interquartile range) & Median (interquartile range) &Median (interquartile range) & Median (interquartile range) \\
\hline
& Fit: total number of deaths & Fit: total hospitalizations & Fit: $u,v$ deaths separately & Fit: $u,v$  hospitalizations separately  \\
& [Model A1, norm Eq.~(\ref{eq:N_DTotal})] & [Model A2, norm Eq.~(\ref{eq:N_JTotal}) ]
& [Model A1, norm Eq.~(\ref{eq:N_DuDv})] & [Model A2, norm Eq.~(\ref{eq:N_JuJv})] \\
\hline
$\beta^{uu}$ & 0.1682 (0.1491--0.1915) & 0.2563 (0.2241--0.2855) & 0.1607 (0.1339--0.1862) & 0.3418 (0.3300--0.3539)\\
$\beta^{uv}$ & 0.2450 (0.2088--0.2840) & 0.2084 (0.1798--0.2328) & 0.2023 (0.1840--0.2235) & 0.1509 (0.1216--0.1846)\\
$\beta^{vu}$ & 0.0780 (0.0541--0.0979) & 0.0709 (0.0357--0.0995) & 0.0261 (0.0158--0.0399) & 0.0500 (0.0453--0.0551)\\
$\beta^{vv}$ & 0.0977 (0.0790--0.1144) & 0.1422 (0.1123--0.1711) & 0.2266 (0.1917--0.2625) & 0.1885 (0.1737--0.2037)\\
$1/\sigma_1$ & 1.7558 (1.6627--1.8545) & 1.7605 (1.6190--1.9650) & 1.7811 (1.6837--1.8945) & 1.7169 (1.6315--1.8140)\\
$1/\sigma_2$ & 1.6798 (1.5660--1.8040) & 1.6690 (1.4572--1.8302) & 1.6384 (1.5075--1.7495) & 1.7085 (1.5905--1.8255)\\
$\phi_u$ & 0.2469 (0.2333--0.2588) & 0.2659 (0.2472--0.2865) & 0.3125 (0.2885--0.3366) & 0.3360 (0.2880--0.3803)\\
$\phi_v$ & 0.2551 (0.2465--0.2617) & 0.2680 (0.2546--0.2834) & 0.3346 (0.3063--0.3535) & 0.3161 (0.2936--0.3337)\\
$1/\mu_u$ & 3.1794 (3.0941--3.2628) & 3.1836 (3.0829--3.3157) & 3.2599 (3.1700--3.4231) & 3.4953 (3.2164--3.9040)\\
$\gamma_{r,u}$ & 0.1509 (0.1335--0.1661) & 0.1845 (0.1678--0.2006) & 0.1386 (0.1263--0.1511) & 0.1731 (0.1562--0.1867)\\
$\gamma_{h,u}$ & 0.0115 (0.0091--0.0133) & 0.0037 (0.0034--0.0042) & 0.0107 (0.0093--0.0121) & 0.0039 (0.0038--0.0041)\\
$1/\mu_v$ & 3.2920 (3.2141--3.4560) & 3.3012 (3.1761--3.4442) & 3.0606 (2.8636--3.2685) & 3.2347 (3.0419--3.4284)\\
$\gamma_{r,v}$ & 0.1607 (0.1431--0.1707) & 0.1679 (0.1357--0.1892) & 0.1627 (0.1469--0.1892) & 0.1424 (0.1315--0.1571)\\
$\gamma_{h,v}$ & 0.0106 (0.0095--0.0120) & 0.0028 (0.0025--0.0031) & 0.0105 (0.0095--0.0118) & 0.0029 (0.0028--0.0029)\\
$\kappa_{r,u}$ & 0.2521 (0.2216--0.2852) & --- & 0.1882 (0.1515--0.2290) & ---\\
$\kappa_{d,u}$ & 0.0185 (0.0162--0.0207) & --- & 0.0186 (0.0178--0.0193) & ---\\
$\kappa_{r,v}$ & 0.1886 (0.1314--0.2738) & --- & 0.1688 (0.1576--0.1808) & ---\\
$\kappa_{d,v}$ & 0.0071 (0.0060--0.0084) & --- & 0.0109 (0.0102--0.0116) & ---\\
$\theta$ & 0.8483 (0.8417--0.8552) & 0.8662 (0.8529--0.8871) & 1 & 1\\
\hline
Initial condition && & & \\
\hline
$E_u(0)$ & 5648 & 8234 & 2690 & 6483 \\
$P_u(0)$ & 4767 & 4340 & 2677 & 5530 \\
$A_u(0)$ & 3609 & 3762 & 988 & 3110 \\
$I_u(0)$ & 11318 & 12759 & 5189 & 12655 \\
$H_u(0)$ & 295 & --- & 278 & --- \\
$E_v(0)$ & 9401 & 8642 & 10330 & 6477 \\
$P_v(0)$ & 7936 & 4555 & 10278 & 5524 \\
$A_v(0)$ & 6007 & 3949 & 3795 & 3107 \\
$I_v(0)$ & 18839 & 13392 & 19925 & 12644 \\
$H_v(0)$ & 381 & --- & 200 & --- \\
\hline \hline
\end{tabular}  \end{center}
\end{table}

\subsection{Model A2: Branches terminate at hospitalizations}
\label{sec:ModelA2}

\begin{figure}[htb]
    \centering
    \includegraphics[width=0.5\textwidth]{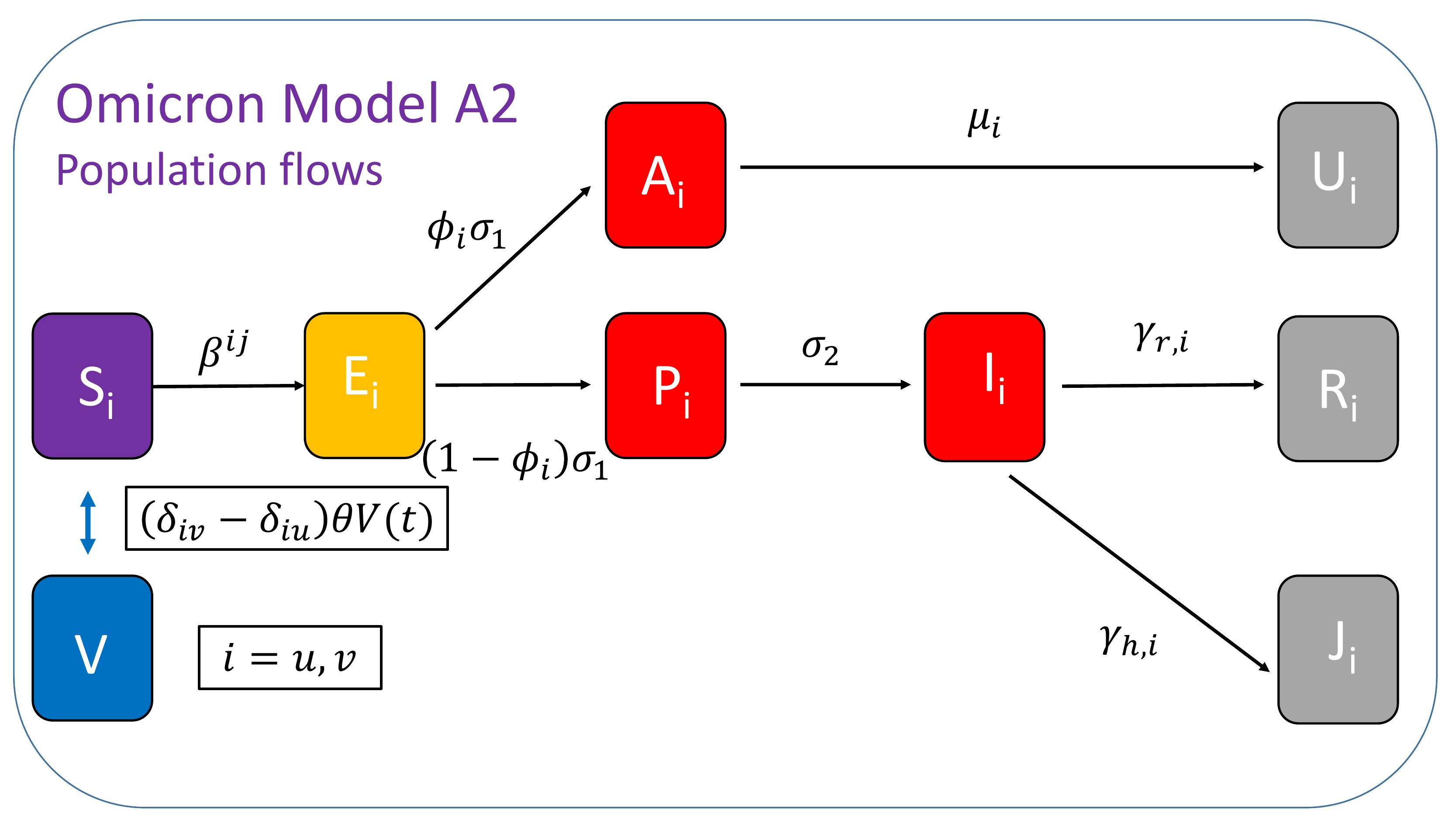}
    \caption{Schematic diagram of the population flows according to model A2.
    The susceptible interactions are as in model A1, shown in the the right panel of Fig.~\ref{fig:ModelA1}.
    The symbol $\delta_{ij}$ with $i,j = u,v$ is the Kronecker delta.}
    \label{fig:ModelA2}
\end{figure}

We now consider model A2, an omicron-variant model similar to A1, but
where the population branches terminate at the total number of hospitalizations.
The total hospitalization data appear, in our gauge, to be more reliable than fatalities, as the latter (at least in Andalusia)
include deceased by any cause that may have recently
generated a positive test. That is to say, we believe that
numerous fatalities were attributed to COVID even though
the primary reason for these events had not been COVID, but
another occurrence, see, for example,~\cite{deathsVsHosp}.
By considering the reported (total,  conventional and critical)
hospitalizations, this possible misattribution of
fatalities to COVID-19 may be diminished.

Model A2 is the same as model A1 described by Eqs.~(\ref{ceq1}),  differing only
in the terminal compartments of hospitalizations. This implies that the
ODEs Eqs.(\ref{Su1})--(\ref{Iu1}) and
Eqs.~(\ref{Sv1})--(\ref{Iv1}) form part of
the model A2 equations, as well. However,
Eqs.~(\ref{Hu1})--(\ref{Du1})
and Eqs.~(\ref{Hv1})--(\ref{Dv1})
are to be replaced by
\begin{equation}
\begin{split}
\frac{\mathrm{d}J_i}{\mathrm{d}t} &= \gamma_{h,i}I_i, \quad i = u,v, \\
\frac{\mathrm{d}R_i}{\mathrm{d}t} &= \gamma_{r,i}I_i, \quad i = u,v
\end{split}
\end{equation}
where $J(t)$ in the total number of hospitalizations and $\gamma_{ij}$ is the rate symptomatically
infected individuals  $u,v$  become hospitalized $\gamma_{h,i}$ or recover $\gamma_{r,i}$.

The optimal parameters (and initial conditions) are obtained
by a procedure similar to what we used in model A1 with $J(t)$ playing the role of $D(t)$.
Accordingly, the norms change: Eq.~(\ref{eq:N_DTotal}) becomes
\begin{equation}\label{eq:N_JTotal}
{\mathcal N}=\frac{1}{n}\sum_{i=1}^n\Big\{\log\big[J_{u,\mathrm{num}}(t_i)+J_{v,\mathrm{num}}(t_i)\big]-\log\big[J_{\mathrm{obs}}(t_i)\big]\Big\}^2 , \end{equation}
while Eq.~(\ref{eq:N_DuDv}) becomes
\begin{equation}\label{eq:N_JuJv}
     {\mathcal N}=\frac{1}{n}\sum_{i=1}^n \Big \{ \log \big [ J_{u,\mathrm{num}}(t_i)\big ] -
     \log \big [ J_{u,\mathrm{obs}}(t_i) \big ] \Big \}^2
    +\Big \{ \log \big [ J_{v,\mathrm{num}}(t_i) \big ]
    - \log \big [ J_{v,\mathrm{obs}}(t_i) \big ]  \Big \}^2 .
\end{equation}

\subsubsection{Andalusia}
\label{sec:AndalusiaA2}

\begin{figure}[htb]
    \centering
    \includegraphics[width=0.5\textwidth]{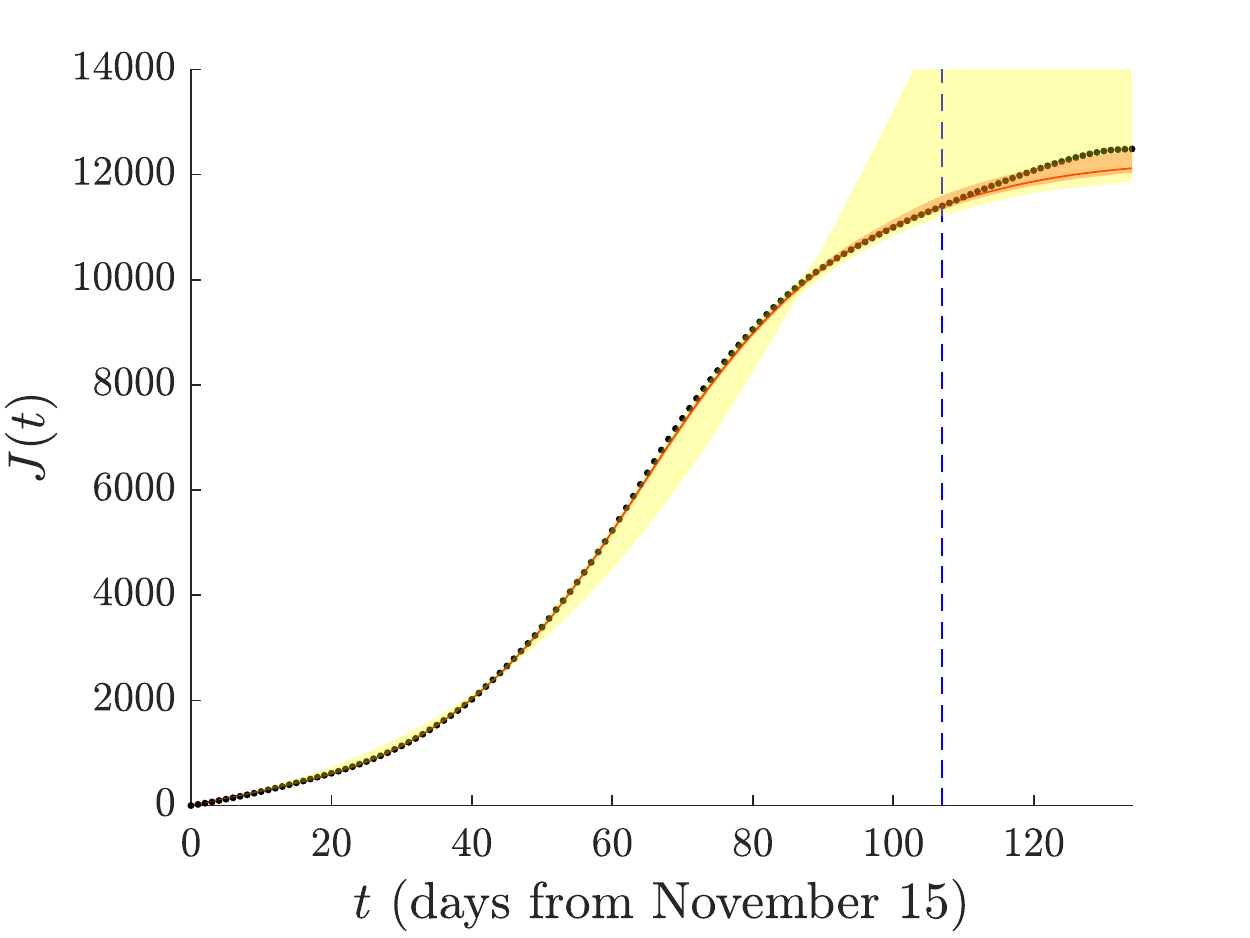}
     \caption{Omicron-variant model A2: Fit and prediction for the total number
    of hospitalizations in Andalusia (norm  of Eq.~(\ref{eq:N_JTotal})). The calculated curve is plotted in red along
    with its confidence/prediction intervals: red shade corresponds to the
    interquartile range, yellow shade to the 95\% confidence interval comprised between the 2.5 and 97.5 percentiles.
    Reported data for the total number of hospitalizations are given by the black points.
    The vertical line, the beginning of the prediction interval, is March 1, 2022.}
    \label{fig:total_hosp}
\end{figure}

The results of fitting the total hospitalizations
(arising from both vaccinated plus unvaccinated populations
using norm~(\ref{eq:N_JTotal})) are shown in Figure \ref{fig:total_hosp}.
The optimal parameters and initial conditions are summarized in Table \ref{tab:andalusia}, last column.

As in the case of the cumulative optimization of
both the vaccinated and the unvaccinated fatalities,
the results of Fig.~\ref{fig:total_hosp} appear quite accurate,
including the forward prediction for the month of
March (despite a slight under-prediction of hospitalizations).
Note that model A1 predictions were slightly above reported fatalities.
However, a more careful inspection of the obtained parameters
suggests that some of them may not be epidemiologically realistic.
Inspection of the optimized transmission rates
shows that the $vv$ rate is larger than the $uu$ rate ($\beta^{vv} > \beta^{uu}$
and $\beta^{vu} > \beta^{uv}$ and $> \beta^{uu}$).
Whereas these inequalities may be related to changes in the
behavior of vaccinated individuals, (for example, vaccinated individuals
may take fewer precautions and socialize more) we believe instead that
this aspect points to the non-identifiability of the model.
It is also likely that the origin of these transmission rates
stems from
our regression's inability to expressly
distinguish between the two $u,v$ compartments,.
More concretely, similarly to model A1, the appropriate identifiability analysis should consider that
the only output is the cumulative number of hospitalizations $J_u(t)+ J_v(t)$.
However, for such an output we could not obtain any identifiability results following
the procedure described in Appendix~\ref{sec:IA}. The model is too complex and beyond the capability of the current
identifiability-analysis packages.

\subsubsection{Switzerland}
\label{sec:SwitzerlandA2}

We followed the same procedure as that for Andalusia to generate the
model A2 fits and predictions for Switzerland (including the Principality of Liechtenstein).
As for model A1, we fitted
both the total number of hospitalizations via  norm~(\ref{eq:N_JTotal})
and the two vaccination-identified compartments via norm~(\ref{eq:N_JuJv}).
We followed the same procedure
as that used to determine $D_u(t)$ and $D_v(t)$ to obtain estimates
for $J_u(t)$ and $J_v(t)$.

\begin{figure}[htb]
    \centering
    \includegraphics[width=0.5\textwidth]{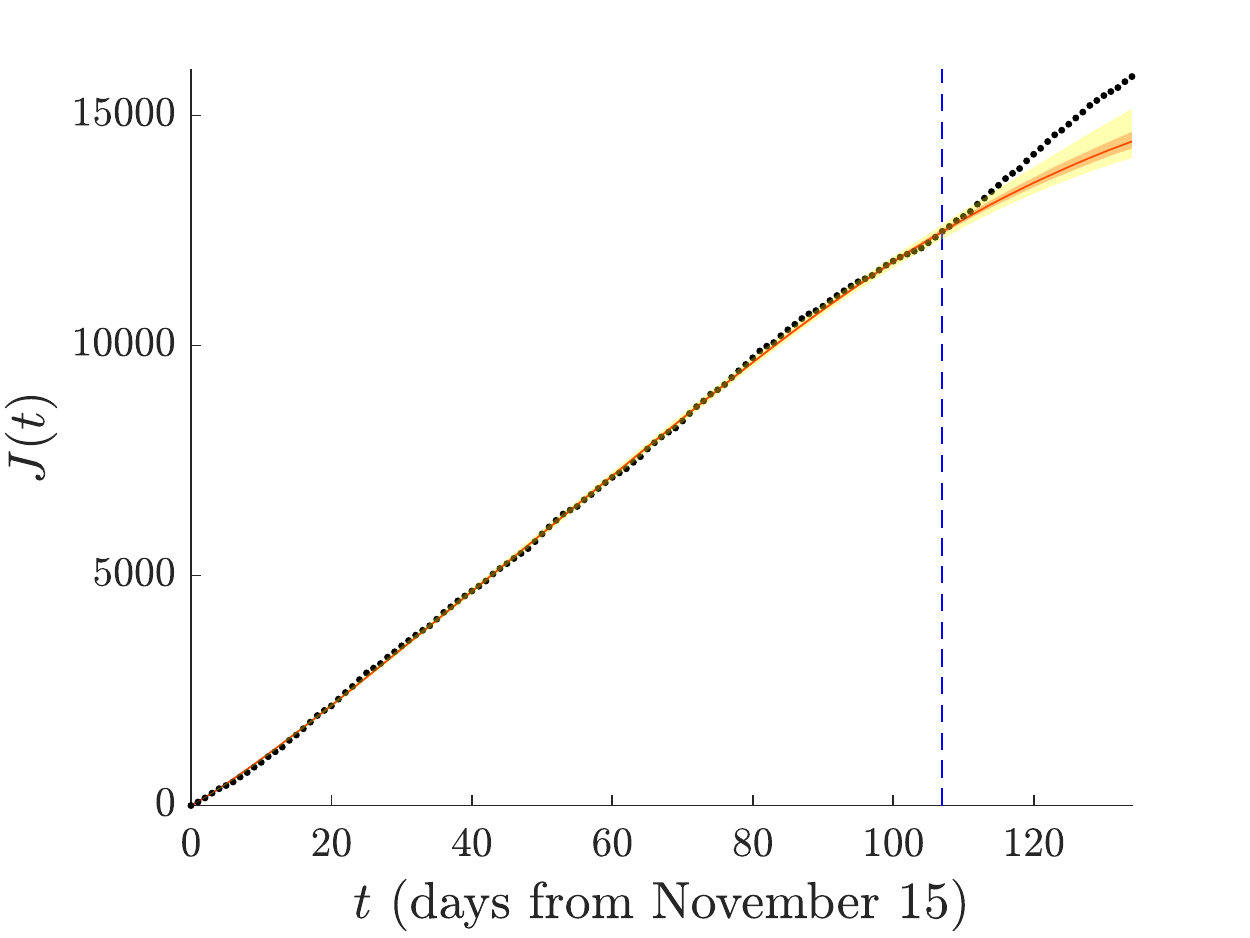}
    \caption{Omicron-variant model A2: Fit and prediction of the  total number of hospitalizations
     in Switzerland (norm  of Eq.~(\ref{eq:N_DTotal})).
     The calculated curve is plotted in red along
    with its confidence/prediction intervals: red shade corresponds to the
    interquartile range, whereas the yellow shade presents the 95\% confidence interval comprised between the 2.5 and 97.5 percentiles.
    Reported data for the total number of hospitalizations are given by the black points.
    The vertical line, the beginning of the prediction interval, is March 1, 2022.}
    \label{fig:total_hosp_chfl}
\end{figure}

\begin{figure}[htb]
    \begin{center}
    \begin{tabular}{cc}
    \includegraphics[width=0.5\textwidth]{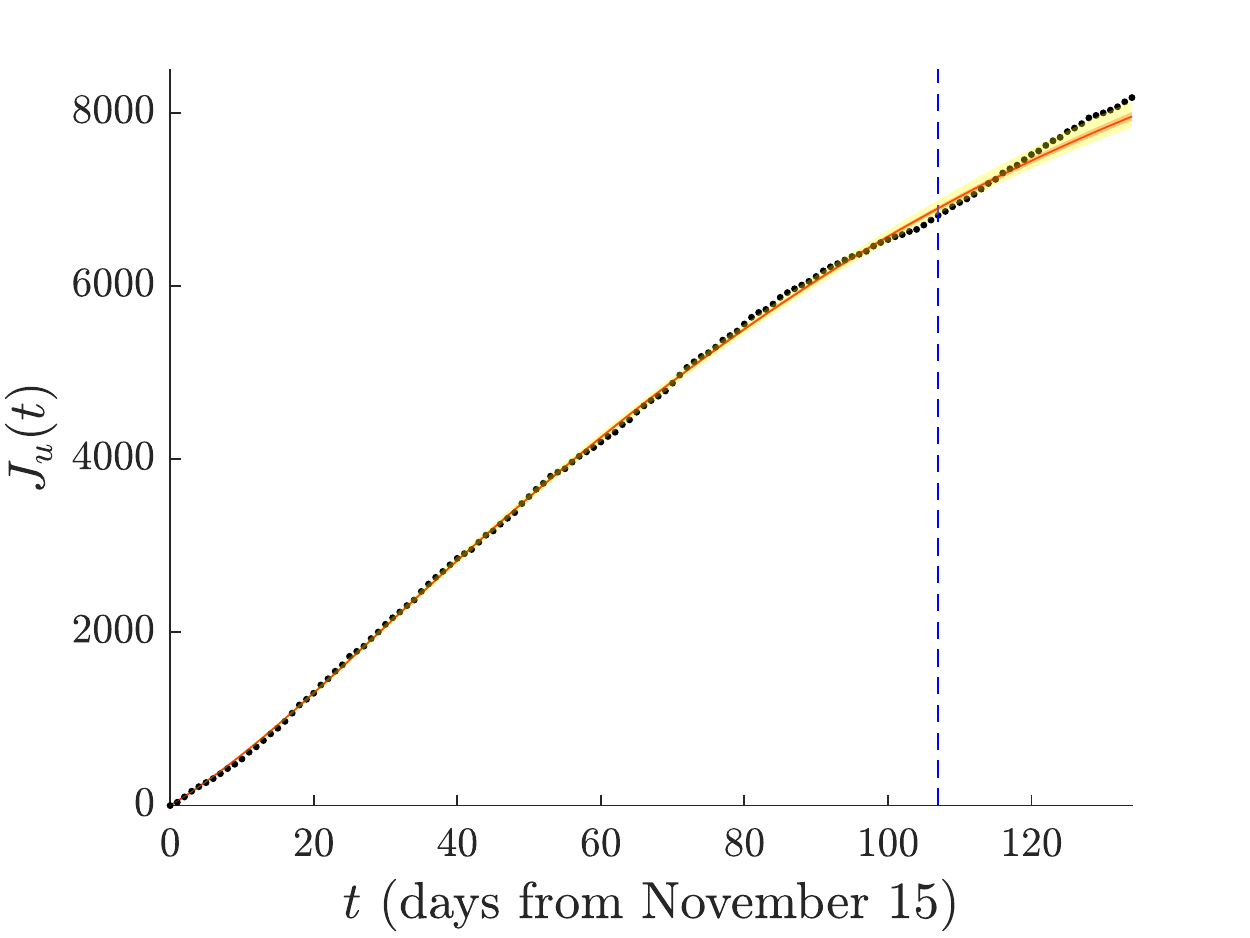} &
    \includegraphics[width=0.5\textwidth]{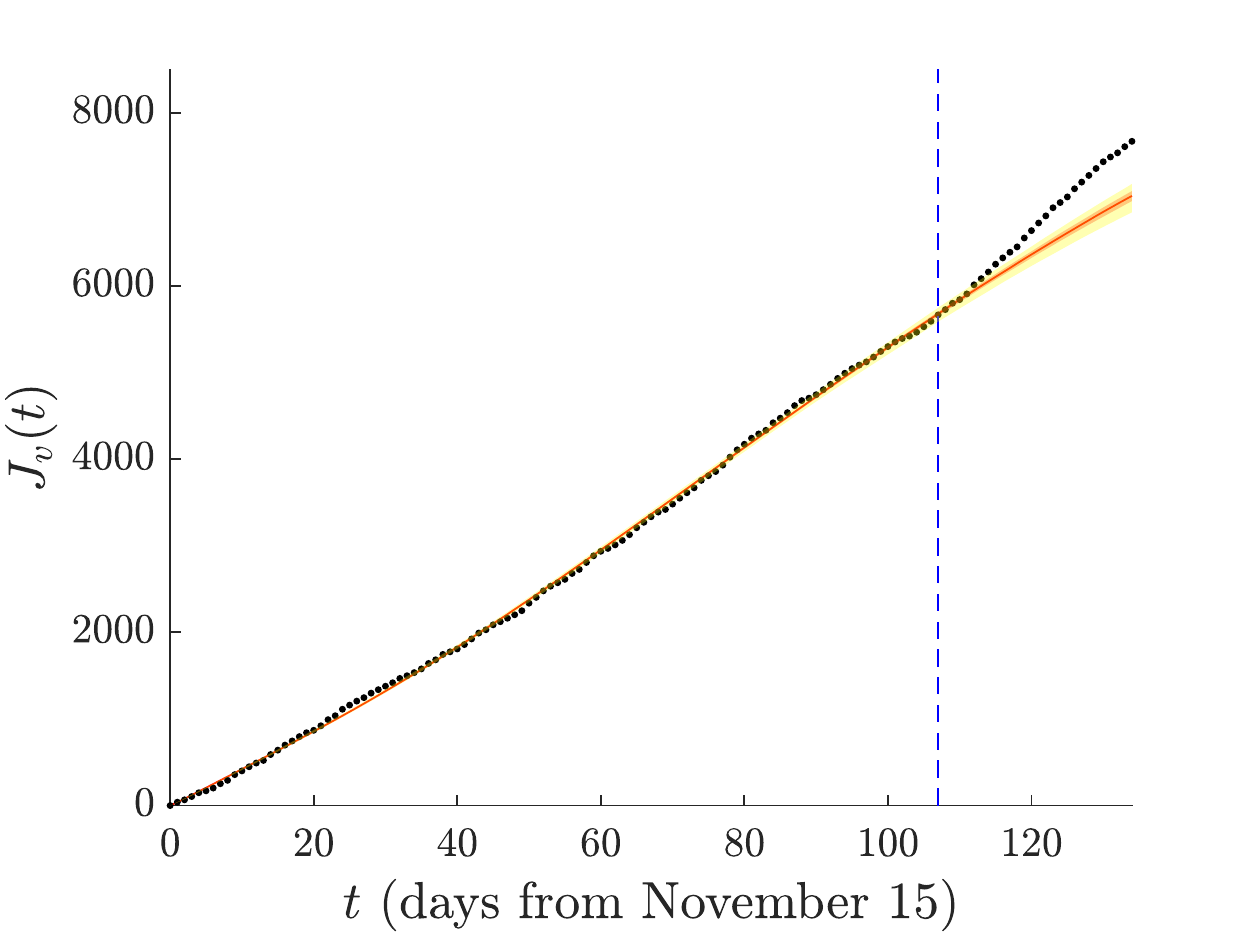} \\
    \end{tabular}
    \end{center}
     \caption{Omicron-variant model A2: Separate fits of vaccinated and
     unvaccinated hospitalizations in Switzerland
    (norm  Eq.~(\ref{eq:N_DuDv})).  Left panel: Hospitalizations of vaccinated individuals.
    Right panel: Hospitalizations of unvaccinated individuals.
    Calculated curves are plotted in red along
    with its confidence/prediction intervals: red shade corresponds to the
    interquartile range, whereas the yellow shade presents the 95\% confidence interval comprised between the 2.5 and 97.5 percentiles.
    Reported data for the total number of vaccinated and unvaccinated hospitalizations are given by the black points.
    The vertical line, the beginning of the prediction interval, is March 1, 2022.}
    \label{fig:hosp_chfl}
\end{figure}

Figure~\ref{fig:total_hosp_chfl} presents our results
for the fit to total number of hospitalizations.
Figure~ \ref{fig:hosp_chfl}, instead,  corresponds to the vaccinated
and unvaccinated populations considered separately in the regression. Table~\ref{tab:switzerland},
third and fifth column,
summarizes  all the fitting parameters and initial conditions.

We can see a  clear model under-prediction of hospitalizations,
despite an accurate following of the time-series throughout
the period over which regression is performed.  The under-prediction is
more pronounced in the case of the fit to the
total number of hospitalizations. In the case of the
separate fittings, the hospitalizations of the vaccinated population
are more under-predicted than the hospitalizations
of unvaccinated individuals.
We attribute this
to the fact that, as also discussed in the context of
fatalities, towards the end of the fitting period,
restrictions were considerably relaxed leading to more cases,
and eventually more fatalities.  As regards the parameters of the model,
we find the transmission rates to be more in line with what
one might typically expect. In particular,
$\beta^{uu} > \beta^{vv}$ for both fittings (with the two different norms),
but $\beta^{uv} > \beta^{vv}$ for the fitting to the total hospitalization,
whereas the reverse
is true for the fitting to the two separate populations (although
in both cases, the rates are fairly similar, even more so when
considering the interquartile ranges).
The fraction of asymptomatics varies  from approximately 27\% to 33\%,
again in reasonable agreement with~\cite{AsymptomaticsReview2022}.
The remaining parameters do not seem to
significantly depend on the norm chosen.

In our identifiability analysis of model A2 with the two vaccination
compartments treated separately we considered, as in the case of model A1,
that $J_u(t)$ and $J_v(t)$ are separately, and continuously,  known.
Moreover, we took  $\theta=1$. In the case of model A2 and
under the above assumptions, following the procedure
described in Appendix~\ref{sec:IA}, we can show
that all  parameters and initial conditions are globally identifiable. This implies that,  {\it in principle}, all
initial conditions and parameters can be determined from the output $J_u(t)$ and $J_v(t)$.
Practically speaking, however,
only the discrete time series $J_u(t)$ and $J_v(t)$ are known: we do
not have the full information of the continuous changes of $J_u(t)$ and $J_v(t)$ as the identifiability
analysis supposes. Hence,  the
loss function used in the parameter estimation is
based on a discrete time series reflecting a finite number of observations.
Given the complexity of the model and the large number of parameters involved, the optimization package often fails to find a minimum.
To alleviate the situation, we chose to fix certain initial conditions, even though
such a choice is inconsistent
with being globally identifiable.
Combined with the possible sloppiness of the model,
the result of such a choice may be that the estimates
for some of the parameters may not be as sharp.
We indicate the above to mitigate a potential impression (to the
reader) that the mathematically obtained global identifiability
of the model should be expected to translate into the most
definitive model results.

\section{Delta variant}
\label{sec:Delta}

Having explored the more elaborate model setting of the omicron variant
we now turn to the simpler case of the delta variant.
What simplifies the model considerably is that it is sufficient to consider
a single susceptibles population since infection of vaccinated individuals
was rare.  Consequently, susceptibles who are vaccinated are added to a ``withdrawn''
population. An alternative option is to add them to the recovered population
in the sense that this is a \emph{terminal} compartment of the model.
For the time period of 3-4 months for which the delta variant
was dominant, the potential waning of immunity
(either from recovery or from the vaccine)
is not considered sufficient to allow these
individuals to replenish the susceptibles compartment. For
more persistent variants, replenishing the susceptibles
population may be relevant.
\begin{figure}[htb]
    \centering
    \includegraphics[width=0.45\textwidth]{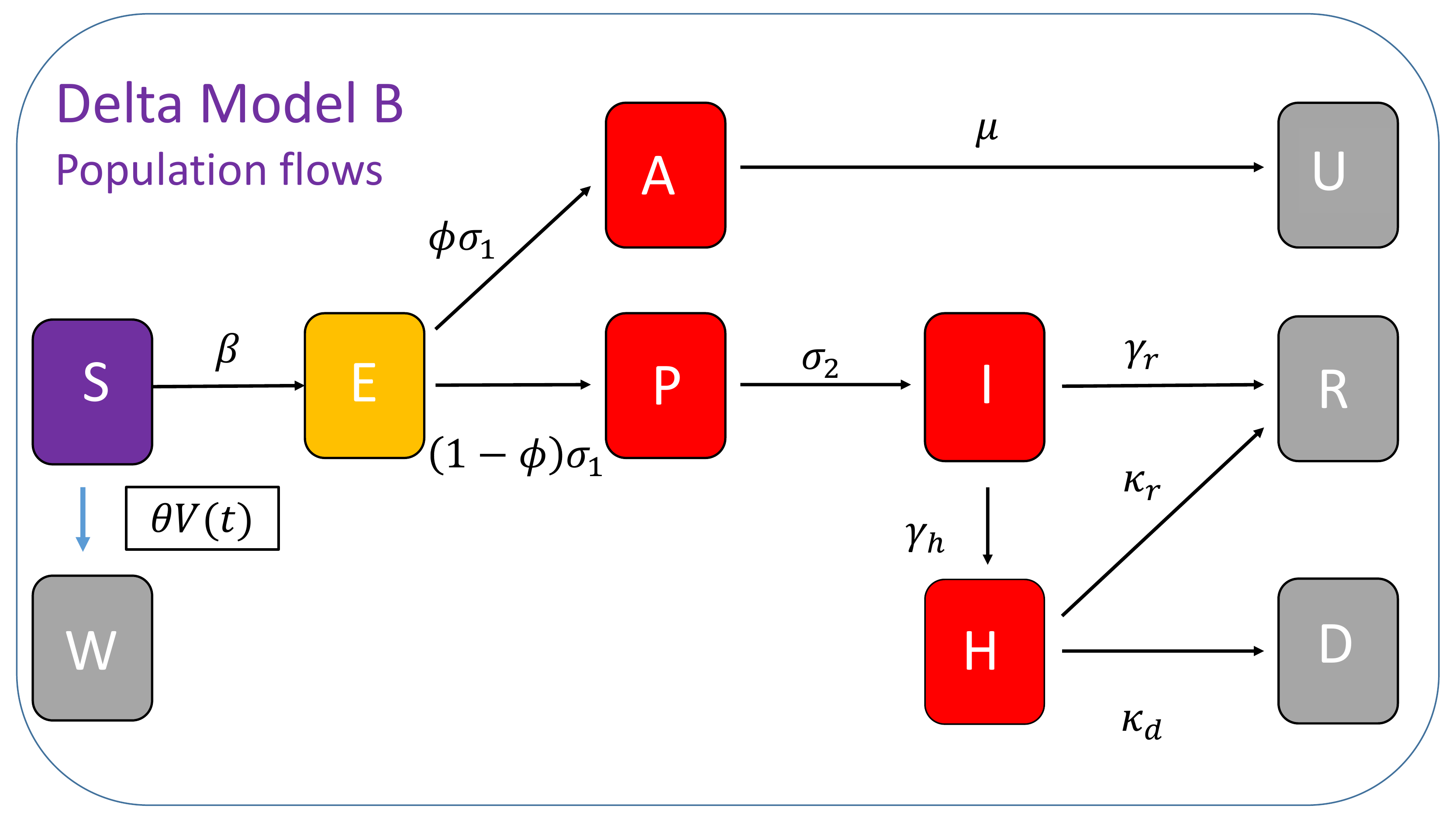}
    \hspace{1.5cm}
    \includegraphics[width=0.45\textwidth]{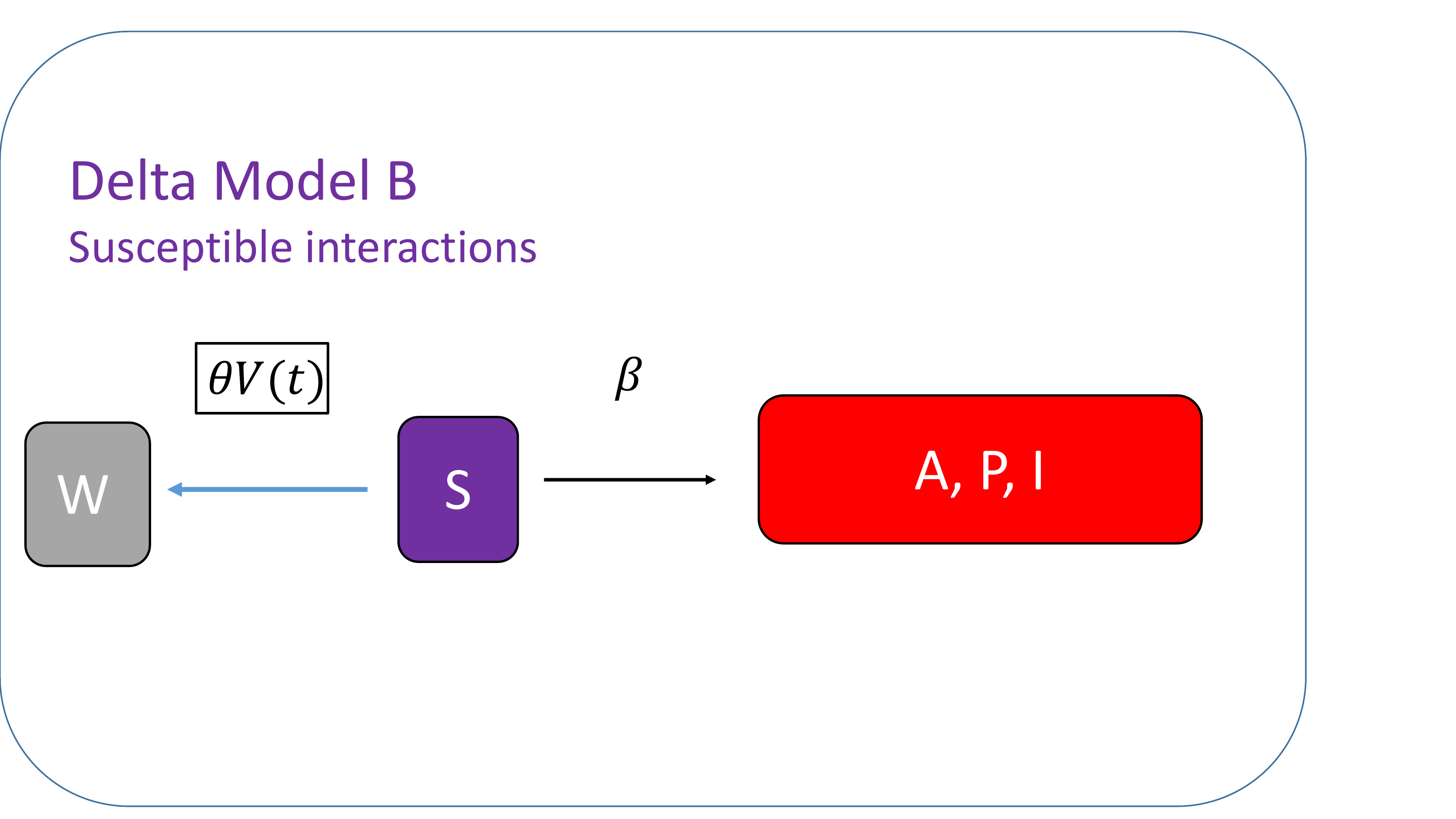}
    \caption{Schematic diagram of population flows according
    to the delta-variant model B (left panel) and susceptible interactions with other population
    compartments (right panel).  A single population is modeled, as we consider that the waning immunity time scale
    (either due to vaccine immunity or due to recovery) is much longer than
    the time scale of prevalence of the delta variant. Vaccinated individuals
    W are permanently removed from the susceptible compartment, their population
    becoming a terminal compartment of the model.}
    \label{fig:ModelB}
\end{figure}
Furthermore,  the SARS-CoV-2 vaccines were highly effective against
the delta variant, leading to rather few breakthrough
infections. This fact removes the need for a detailed
modeling of compartments within the vaccinated category.
Accordingly, the relevant model with the same (but single-component)
populations as before and with the addition of the withdrawn ($W$)
compartment reads:
\begin{equation}
\begin{split}
\frac{\mathrm{d}S}{\mathrm{d}t} &= -\beta S(I+A+P)-\theta V(t) , \\
\frac{\mathrm{d}E}{\mathrm{d}t} &= -\sigma_1E+\beta S(I+A+P) , \\
\frac{\mathrm{d}P}{\mathrm{d}t} &= (1-\phi)\sigma_1E-\sigma_2P , \\
\frac{\mathrm{d}A}{\mathrm{d}t} &= \phi\sigma_1E-\mu A , \\
\frac{\mathrm{d}U}{\mathrm{d}t} &= \mu A , \\
\frac{\mathrm{d}I}{\mathrm{d}t} &= \sigma_2P-(\gamma_r+\gamma_h)I , \\
\frac{\mathrm{d}H}{\mathrm{d}t} &= \gamma_h I-(\kappa_r+\kappa_d) H , \\
\frac{\mathrm{d}R}{\mathrm{d}t} &= \gamma_r I+\kappa_r H , \\
\frac{\mathrm{d}D}{\mathrm{d}t} &= \kappa_d H , \\
\frac{\mathrm{d}W}{\mathrm{d}t} &= \theta V(t) .
\end{split}
\end{equation}
A schematic of the population flows (left panel) and the susceptible
interactions is shown in Fig.~\ref{fig:ModelB}.

The initial conditions are taken in a similar fashion as in the omicron variant,
except
for $S(0)$, which is taken as a fitting parameter.
We chose to render it a fitting parameter
since susceptibles who became infected with previous variants
are immune to the delta variant. However,
their number is not definitively known.
Moreover, as in our modeling of the omicron variant via models A1 and A2,
we supposed that the transmission rate $\beta$ is the same for all three
infectious compartments: asymptomatics, presymptomatics, and for the symptomatically infected population.
In addition, as in the case of the omicron-variant models, we imposed
 the single constraint on the incubation period $\tau_{\textrm{inc}} = \sigma_1^{-1}+\sigma_2^{-1}$ to be equal to a value randomly
 sampled following a normal distribution whose mean now is
 4.41 and standard deviation 0.3291, in line with what
 is reported in~\cite{jama2022}.

The norm associated with model B,  and minimized during the optimization
procedure is:
\begin{equation} \label{eq:normB}
    {\mathcal N}=\frac{1}{n}\sum_{i=1}^n \Big \{ \log \big [ D_{\mathrm{num}}(t_i) \big ]
    -\log \big [ D_{\mathrm{obs}}(t_i) \big ] \big \}^2.
\end{equation}

Again, following the approach presented in Appendix~\ref{sec:IA},  we find that the three parameters
 $\sigma_1, \phi, \mu$ and the following combinations
 \begin{eqnarray*}
 (\gamma_h+\gamma_r)+\sigma_2,\quad (\gamma_h+\gamma_r)\cdot \sigma_2,
\quad \kappa_{d}+\kappa_{r}, \quad
\frac{\beta (\gamma_h+\gamma_r)}{\gamma_h \kappa_{d}},  \quad \frac{\gamma_h}{\gamma_h+\gamma_r} \kappa_{d} \theta ,
\end{eqnarray*}
 are globally identifiable. Thus $\sigma_2$ and the sum $(\gamma_h + \gamma_r)$ are locally identifiable.
Initial conditions other than $D(0)$, which is explicitly available through
the fatality time series, are not identifiable.
As discussed in the identifiability
analyses of the two omicron models, this is just the theoretical
result, based on the analysis of~\cite{Pogudin_SIAN}.
As we have empirically observed,
when we fix certain initial conditions or choose a bound for the parameters to be optimized,
the identifiability properties of the model may change.
In that light, the relevant parameter identifications should
be considered with the associated practical ``word of caution''
indicated above.

\subsection{Andalusia and Switzerland}
\label{sec:Results-ModelB}

In our simulations, the fitting window for the Andalusia calculation
started on June 15, 2021, whereas it started on
July 1, 2021 for the Switzerland simulations. {The choice of the initial time was determined
from the existence of a plateau in the associated time series.} The fitting period
ended on October 1, 2021 for both regions, and the prediction
interval terminated on November 1, 2021. As mentioned in Section~\ref{sec:Omicron} the
omicron variant appeared in November 2021.

\begin{figure}
    \centering
    \includegraphics[width=0.45\textwidth]{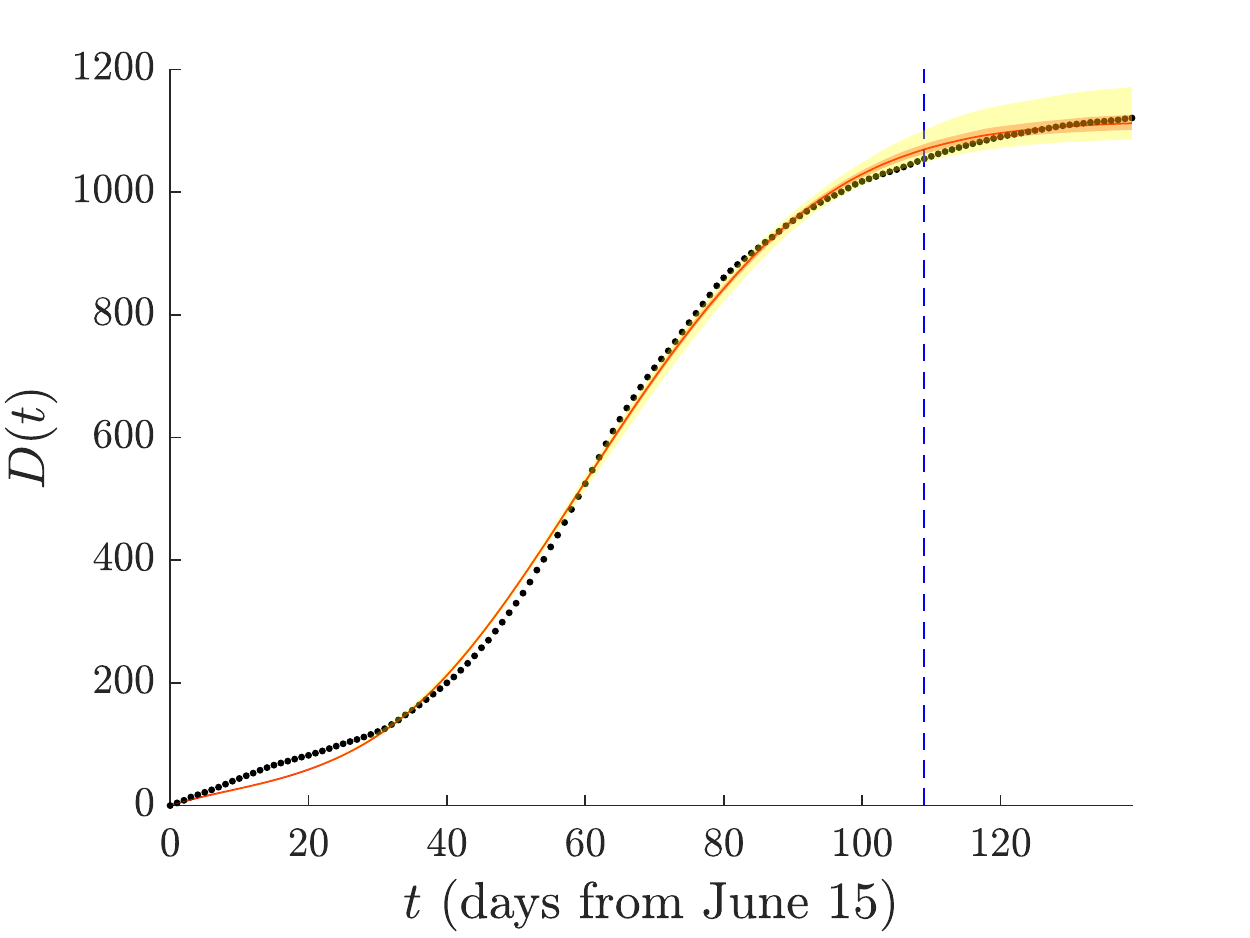}
    \includegraphics[width=0.45\textwidth]{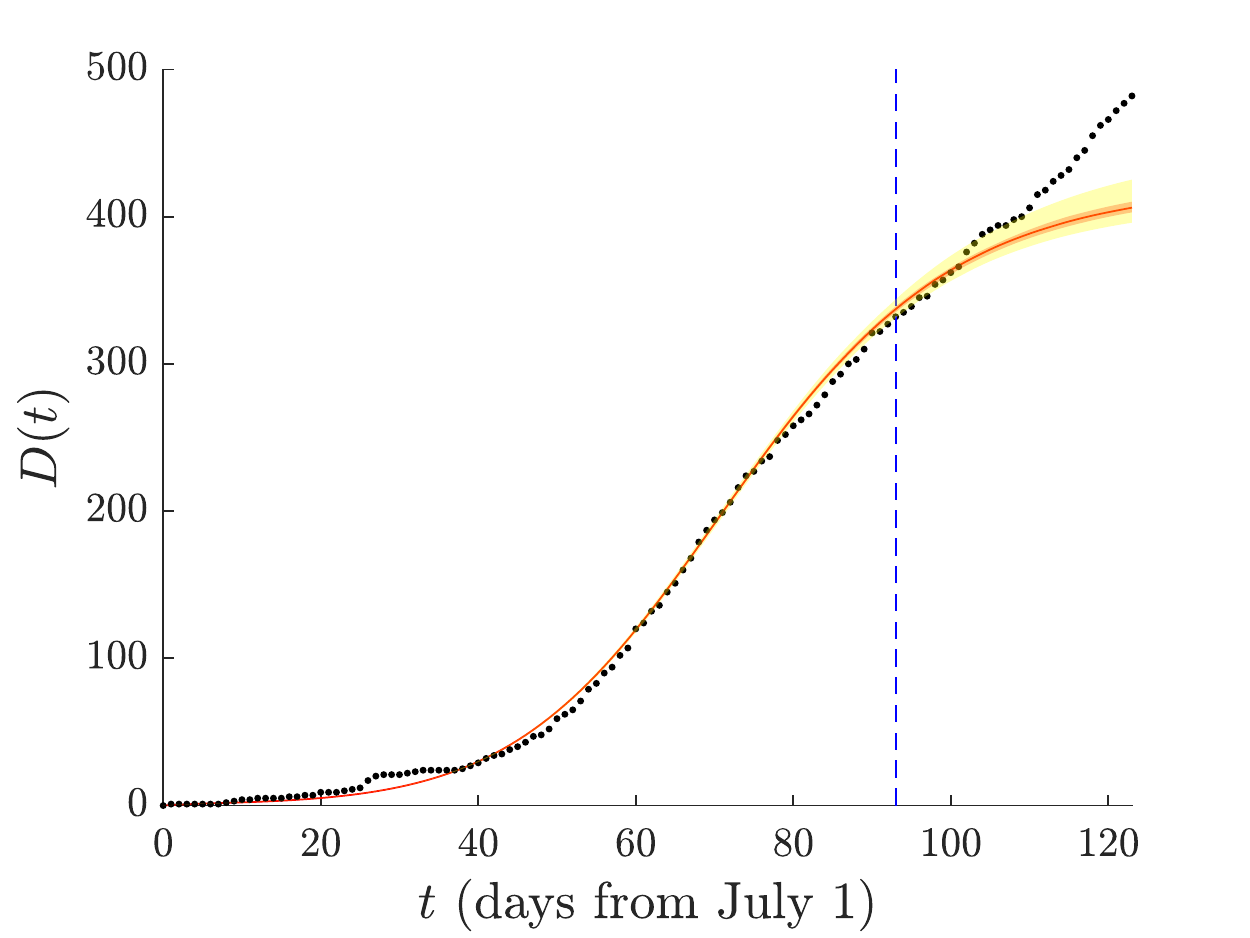}
    \caption{Delta-variant model B: Fit and prediction of the total
    number of fatalities in Andalusia (left panel) and Switzerland (right panel).
    Norm~(\ref{eq:normB}) was used. The vertical line, the beginning of the
    prediction interval, is October 1, 2022. Note the significant
    difference in the number of fatalities.}
    \label{fig:delta_Both}
\end{figure}

Figure \ref{fig:delta_Both} shows the results of our
model B simulations, both for the fitting and the prediction intervals.
The left panel presents results for Andalusia, whereas the right panel for
Switzerland. The best fitting parameters and initial conditions for both
countries  corresponding to the model B ODEs are presented in
Table~\ref{tab:BothModelB}.

In the case
of Andalusia, we observe a high quality fit, not only for the
regression interval but also for the prediction interval.
Nevertheless, some parameters do not seem to
be in agreement with current knowledge of the
epidemiology of the delta variant of SARS-CoV-2. We believe, that the
primary reason is the lack of identifiability of the model
(both the local aspects thereof theoretically, as well as the
practical aspect highlighted above in connection to data and
initial condition choices).
For example, the recovery time of asymptomatics, $1/\mu$ is found
to be $ \approx 3.5$ days, and an upper bound to the
recovery time of symptomatically infected is
$1/\gamma_r \approx 5.4$ days.
It may be expected that both time scales are likely to be longer than
these predictions, although these numbers are in
reasonable correspondence with findings, e.g., such as
the ones of~\cite{10.3389/fimmu.2022.812606} for the delta variant.
On the other hand,  the fraction of asymptomatics, 8\%, is in agreement
with the review and analysis of~\cite{AsymptomaticsReview2022} who found a considerably
smaller fraction of asymptomatics associated with the delta than the omicron
variants, again in agreement with our calculations.

Our model calculations for Switzerland in Fig.~\ref{fig:delta_Both}
provide a reasonable fit throughout the training interval.:
calculations initially under-predict and later over-predict.
Nevertheless, the predicted time series considerably
under-predicts the number of fatalities over the testing
period.  Some optimized
parameters for this territory differ significantly from those obtained
for Andalusia. In particular, the transmission rate for
the Switzerland data is higher than that for the Andalusia data,
as are the $H \rightarrow R$
recovery rates.
The Switzerland parameters, however, for the death rate and the vaccine efficiency are
predicted to be lower than those for Andalusia.
In this case, we do not have a definitive attribution of the relevant result (i.e., the under-prediction of
fatalities) in the case of Switzerland.
The only change in policy that we could identify
was that
from September 13, 2021, access to most indoor public spaces like restaurants, bars, museums or fitness centres was
permitted with a valid COVID certificate in Switzerland.
No other restrictions were enforced on  fully vaccinated and boosted people.

\begin{table}
\begin{center}
\caption{\label{tab:BothModelB}
Optimal parameters and initial conditions for the delta-variant
model, model B, in Andalusia (third column) and Switzerland (fourth column).
Model fits to the total number of fatalities, norm~(\ref{eq:normB}).}
\begin{tabular}{c|c|c|c}
\hline \hline
Parameter & Symbol & Median (interquartile range)
& Median (interquartile range) \\
\hline
& & Andalusia & Switzerland \\
& & Model B, norm~(\ref{eq:normB}) & Model B, norm~(\ref{eq:normB}) \\
\hline
Transmission rate [per day] & $\beta$ & 0.4526 (0.4210--0.4703) & 0.5445 (0.5320--0.5570) \\
Latency period [days] & $1/\sigma_1$ & 2.2001 (2.0896--2.2982) & 2.2027 (2.1113--2.3021) \\
Preclinical period [days] & $1/\sigma_2$ & 2.1590 (2.0518--2.2789) & 2.2056 (2.1112--2.3055) \\
$A/P=$ partitioning [-] & $\phi$ & 0.0801 (0.0780--0.0827) & 0.0800 (0.0787--0.0820) \\
Infectivity period ($A$) [days] & $1/\mu$ & 3.4493 (3.3551--3.5658) & 3.4708 (3.3879--3.6194) \\
Recovery rate $I \rightarrow R$ [per day] & $\gamma_r$ & 0.1852 (0.1643--0.1982) & 0.1888 (0.1842--0.1959) \\
Transition rate $I \rightarrow H$ [per day] & $\gamma_h$ & 0.0017 (0.0016--0.0020) & 0.0026 (0.0024--0.0027) \\
Recovery rate $H \rightarrow R$ [per day] & $\kappa_r$ & 0.0629 (0.0523--0.0831) & 0.2480 (0.2355--0.2637) \\
Death rate $H \rightarrow D$ [per day] & $\kappa_d$ & 0.0462 (0.0449--0.0481) & 0.0069 (0.0066--0.0072) \\
Vaccine efficiency [-] & $\theta$ & 0.7956 (0.7802--0.8240) & 0.6162 (0.6047--0.6189)  \\
Initial ratio [\#] & $S(0)/N(0)$ & 0.5824 (0.5673--0.5951) & 0.5278 (0.5178--0.5377) \\
\hline
Initial conditions & & & \\
\hline
Initial exposed population [\#] & $E(0)$ & 1322 & 669 \\
Initial presymptomatic population [\#] & $P(0)$ & 351 & 591 \\
Initial asymptomatic population [\#] & $A(0)$ & 205 & 332 \\
Initial symptomatically infecgted population [\#] & $I(0)$ & 1914 & 1441 \\
Initial hospitalized population [\#] & $H(0)$ & 80 & 57 \\
\hline \hline
\end{tabular}
\end{center}
\end{table}

\section{Conclusions and Future Challenges}
\label{sec:Conclusions}

In this work we presented a new class of compartmental
epidemiological models
that was motivated by the immunological properties
of the delta and omicron variants of SARS-CoV-2.
More generally, our aim was to present
possibilities for settings where variants are
highly transmissive (and hence relevant to consider
vaccinated individuals and their epidemiological characteristics)
as in the case of models A1-A2 for the omicron variant,
as well as ones where
breakthrough infections are more rare, and hence vaccination
is tantamount to withdrawal from the susceptible population
as in the case of model B for the delta variant.
We, therefore, constructed  model B,
with the stipulation that vaccinated individuals were permanently withdrawn from the susceptible population based on the
vaccination records and vaccine coverage rate.
On the other hand, the epidemiology of the
omicron variant suggests a substantial number of
breakthrough infections,  namely infections of vaccinated
individuals. Accordingly, we developed models for both vaccinated
and unvaccinated populations and analyzed their
pairwise interaction and overall time evolution.
Indeed, two classes of regression results were given.
In the first (and more crude) regression,  only the cumulative number of
fatalities was accounted for in the optimization objective.
This was done when the data did not allow the partitioning
of fatalities (or hospitalizations)
to vaccinated and unvaccinated components.
In the
second, more refined approach,  fatalities (or hospitalizations)
stemming from the
two different (vaccinated or not) groups were separately considered.

We addressed the identifiability of the various models and
considered mathematical issues (e.g.,
parameters globally and locally identifiable, given particular time
series), we raised some practical considerations due to
the finite nature of the available observations, and we
considered the compatibility of the selection of some initial conditions.
In the case where the time series
associated with vaccinated and unvaccinated individuals are required,
we identified the issue of how
to handle the so-called
``unknown'' deaths if the individual vaccination status remains undeclared.
We proposed a concrete approach to address such disparities, yet clearly
these topics merit further investigation.

In our presentation, we focused on
the region of Andalusia in Spain and the country of Switzerland
(which included data from the Principality of Liechtenstein).
These two territories have similar populations. In each territory,
we presented studies 
of a regression effort involving the fatalities (model A1), as well as
one terminating at the compartment of total (i.e., conventional plus critical) hospitalizations (model A2).
Our models gave generally good agreement with the corresponding
training sets, but also reasonable prediction intervals
in comparison with the testing data for periods of about a month
beyond the end of the training period (up to which the optimization is performed).
In the cases where deviations from the predictions were more
significant, plausible explanations were offered
on the basis of, e.g., the relaxation
of measures or other changes of policies.

Naturally, these models offer a starting point for further
considerations and are intended as a stepping stone for
further studies. On the one hand, it would be quite relevant
to seek additional sources of data and other approaches
to parameter estimation (than the regression and bootstrapping
methodologies
used here),  to incorporate more accurately the measurement
uncertainty and to improve the adequacy of the parameter estimation,
in line with our expectations stemming from the analysis
of the model identifiability. Another
important direction is to add the spatial dimension to the proposed
well-mixed ODE models, to incorporate the mobility of
vaccinated individuals. This can be done either at the level of
metapopulation models~\cite{review_meta,vespi2008,rapti2022} or
at that of PDE approaches~\cite{mammeri2020,viguerie2021,kevrekidis2021,theo}.
Finally, numerous additional dimensions of such modeling of vaccinations
are relevant to consider such as, e.g., the age stratification of such effects~\cite{hethcote,agecovid,cuevas2021}.
These directions are currently under consideration and will be reported in future publications.

\section{Declaration of competing interest}

The authors declare that they have no known competing financial
interests or personal relationships that could have appeared to influence
the work reported in this paper.

\section{Acknowledgements}
JC-M acknowledges support from EU (FEDER program 2014-2020) through both Consejería de Economía, Conocimiento, Empresas y Universidad de la Junta de Andalucía (under the projects P18-RT-3480 and US-1380977), and MCIN/AEI/10.13039/501100011033 (under the project PID2020-112620GB-I00). PGK and GAK acknowledge support through the C3.ai Inc. and Microsoft Corporation. The authors thank Gleb Pogudin for his invaluable help in the identifiability analysis work.

\appendix
\section{Identifiability Analysis} \label{sec:IA}

Following the differential algebraic approach in~\cite{Eisenberg2013}, one needs to rewrite the whole system as a single high-order differential
equation of the observable (i.e., the data). Inevitably, a moderate system leads to an equation with a huge number of terms which is very likely
beyond the capability of symbolic mathematics software like Mathematica. The difficulty is due to the nonlinear terms in the original system.
Our approach is to leave some original equation(s) untouched and rewrite the rest as a high-order differential equation of the observable.
Basically, we explicitly carry out as many derivations as possible, and stop short of writing the original system into a single ODE, as the last step(s) may
lead to an exceedingly complex equation. Then we apply the identifiability analysis package \textit{SIAN}~\cite{Pogudin_SIAN} and \textit{StructuralIdentifiability}~\cite{Pogudin_2022, Pogudin_2021} to the new system.
In what follows, we explain how it is done for the model given by Eqs. ~(\ref{ceq1}), model A1. The other two models, models A2 and B,
are analyzed in the exactly same way. Note that this approach
is not limited to the models presented in this work. In addition,
we remark that results could not be obtained for all the cases.

The subscript $u$ or $v$ will be dropped whenever there is no obvious ambiguity. Introduce intermediate parameters
$\gamma_s = \gamma_r + \gamma_h$ and $\kappa_s = \kappa_r + \kappa_d$.
The idea is to rewrite Eqs.~(\ref{ceq1}) as a system of equations
for the number of asymptomatics $A$ and a high-order ODE for the number of fatalities $D$'s.
The $D$'s need to be kept
because they are the output (i.e., the observable).

For the unvaccinated state variables,  Eqs.~(\ref{Du1}, \ref{Hu1}, \ref{Iu1}, \ref{Pu1}) lead to
\begin{eqnarray*}
H&=& \frac{1}{\kappa_d}D' , \\
I&=& \frac{1}{\gamma_h}(H' + \kappa_s H) = \frac{1}{\gamma_h \kappa_d} (D'' + \kappa_s D') , \\
P&=& \frac{1}{\sigma_2} (I' + \gamma_s I) = \frac{1}{\sigma_2  \gamma_h \kappa_d}( D''' + D'' (\kappa_s +\gamma_s) + D' (\gamma_s \kappa_s) ), \\
E&=& \frac{P' + \sigma_2 P }{(1-\phi)\sigma_1}= \frac{1}{(1-\phi)\sigma_1\sigma_2 \gamma_h \kappa_d}
\left[D^{(4)} + D'''(\kappa_s + \gamma_s + \sigma_2) + D'' (\kappa_s \gamma_s + \kappa_s \sigma_2
+\gamma_s \sigma_2) + D' (\kappa_s \gamma_s \sigma_2)  \right] .
\end{eqnarray*}
Here all subscripts $u$ are dropped: identical equations for vaccinated state variables should also be considered.
From the last equation, we compute the following (since it resembles certain terms in Eq. ~(\ref{Eu1})),
\begin{equation} \label{eq:E}
E + \sigma_1 \int_{t_0}^t E(\tau) \; d\tau = \frac{1}{(1-\phi)\sigma_1\sigma_2 \gamma_h \kappa_d}
\left[D^{(4)} + k_3 D''' + k_2 D'' + k_1 D'+k_0 D  + \tilde{\alpha} \right] ,
\end{equation}
where $\tilde{\alpha}$ is an integration constant (a new parameter) and
\begin{eqnarray*}
k_3 = \kappa_s + \gamma_s + \sigma_1 + \sigma_2, \quad  k_2 = \kappa_s \gamma_s + \kappa_s \sigma_1 +
\kappa_s \sigma_2 + \gamma_s \sigma_1 + \gamma_s \sigma_2
+ \sigma_1 \sigma_2,\\
 k_1 = \kappa_s \gamma_s \sigma_1 + \kappa_s \gamma_s \sigma_2 + \kappa_s \sigma_1 \sigma_2 + \gamma_s \sigma_1 \sigma_2, \quad
k_0 = \kappa_s \gamma_s \sigma_1 \sigma_2.
\end{eqnarray*}

With Eq.~\eqref{eq:E}, we now add Eqs.~(\ref{Su1}) and (\ref{Eu1}), and then integrate to obtain
\begin{eqnarray*}
(S' + E') = - \sigma_1 E - \theta V(t) \quad \Rightarrow \quad
S = \frac{-1}{(1-\phi)\sigma_1\sigma_2 \gamma_h \kappa_d}
\left[D^{(4)} + k_3 D''' + k_2 D'' + k_1D'+k_0 D  + \alpha + \tilde{\theta} \tilde{V}(t) \right],
\end{eqnarray*}
where $\alpha$ is another integration constant (different from $\tilde{\alpha}$), and
$$
\tilde{V}(t) = \int_{t_0}^t V(\tau) \; d\tau, \quad \tilde{\theta} = (1-\phi)\sigma_1 \sigma_2  \gamma_h \kappa_d \theta .
$$

So far, the state variables for unvaccinated population $H, I, P, E, S$ are expressed as functions of $D$ and its derivatives.
We have identical formulas for the vaccinated populations $H_v, I_v, P_v, E_v, S_v$, except that for $S_v$ a negative sign should
be added in front of $\tilde{\theta}$.

The equation for $E_u$, Eq.~(\ref{Eu1}), multiplied by $(1-\phi_u)\sigma_1\sigma_2  \gamma_{h, u} \kappa_{d, u}$ becomes
\begin{eqnarray*}
D^{(5)}_u &=& -\left[   k_3 D_u^{(4)} + k_2 D_u''' +  k_1 D_u'' +k_0 D_u'  \right] -
\left[D_u^{(4)} + k_3 D_u''' + k_2 D_u'' + k_1D_u'+k_0 D_u  + \alpha_u + \tilde{\theta}_u V(t) \right] \\
& & \qquad \quad \cdot  \left[  \beta^{uu} (I_u + A_u + P_u) +   \beta^{uv} (I_v +A_v+ P_v) \right] .
\end{eqnarray*}
Note that the subscript $u$ for the parameters $k_3, k_2,k_1,k_0$ is still omitted. But for this equation they should be computed from
the parameters associated with the unvaccinated population,
whereas for the equation of $D^{(5)}_v$ they should be computed
from the parameters associated with the vaccinated population.

Furthermore,
\begin{eqnarray*}
\beta^{uu} (I_u + A_u+ P_u ) &=& \beta^{uu} \left[ \frac{D_u'' + \kappa_{s,u} D_u'}{ \gamma_{h, u} \kappa_{d,u}} + A_u +
\frac{D'''_u + D_u''(\kappa_{s,u}+\gamma_{s, u}) + (\kappa_{s,u}\gamma_{s, u}) D_u'  }{\sigma_2  \gamma_{h, u} \kappa_{d,u}}\right] \\
&=&
\frac{\beta^{uu}\left[D_u'' + \kappa_{s,u} D_u'\right]}{ \gamma_{h, u}  \kappa_{d,u} }
+ \frac{\beta^{uu} \left[D'''_u + D''_u(\kappa_{s,u}+\gamma_{s, u}) + (\kappa_{s,u}\gamma_{s, u}) D_u'  \right]}{ \sigma_2 \gamma_{h, u}  \kappa_{d,u} } + \beta^{uu} A_u .
\end{eqnarray*}
Similarly,
\begin{eqnarray*}
\beta^{uv} (I_v + A_v+ P_v ) &=&
\frac{\beta^{uv}\left[D_v'' + \kappa_{s,v} D_v'\right]}{ \gamma_{h, v}  \kappa_{d,v} }
+ \frac{\beta^{uv} \left[D'''_v + D''_v(\kappa_{s,v}+\gamma_{s, v}) + (\kappa_{s,v}\gamma_{s, v}) D_v'  \right]}{ \sigma_2 \gamma_{h, v}  \kappa_{d,v} } + \beta^{uv} A_v  ,\\
\beta^{vu} (I_u + A_u+ P_u ) &=&
\frac{\beta^{vu}\left[D_u'' + \kappa_{s,u} D_u'\right]}{ \gamma_{h, u}  \kappa_{d,u} }
+ \frac{\beta^{vu} \left[D'''_u + D''_u(\kappa_{s,u}+\gamma_{s. u}) +
(\kappa_{s,u}\gamma_{s, u}) D_u'  \right]}{\sigma_2 \gamma_{h, u}  \kappa_{d,u} } + \beta^{vu} A_u , \\
\beta^{vv} (I_v + A_v+ P_v ) &=&
\frac{\beta^{vv}\left[D_v'' + \kappa_{s,v} D_v'\right]}{ \gamma_{h, v}  \kappa_{d,v} }
+ \frac{\beta^{vv} \left[D'''_v + D''_v(\kappa_{s,v}+\gamma_{s, v}) + (\kappa_{s,v} \gamma_{s, v}) D_v'  \right]}{\sigma_2 \gamma_{h, v}  \kappa_{d,v} } + \beta^{vv} A_v.
\end{eqnarray*}
The next step is to scale variables as follows
\begin{equation}
\gamma_{h, u} \kappa_{d, u} A_u \to A_u, \gamma_{h, v }\kappa_{d, v} A_v \to A_v, \quad
\frac{\beta^{uu}}{\gamma_{h, u} \kappa_{d,u}} \to \beta^{uu}, \frac{\beta^{uv}}{\gamma_{h, v} \kappa_{d,v}} \to \beta^{uv},
 \frac{\beta^{vu}}{\gamma_{h, u} \kappa_{d, u}} \to \beta^{vu},  \frac{\beta^{vv}}{\gamma_{h, v} \kappa_{d,v}} \to \beta^{vv} .
\end{equation}

The above equation for $D^{(5)}_u$ becomes
\begin{eqnarray*}
D^{(5)}_u &=&-\left[   k_3 D_u^{(4)} + k_2 D_u''' +  k_1 D_u'' +k_0 D_u'  \right] -
\left[D_u^{(4)} + k_3 D_u''' + k_2 D_u'' + k_1D_u'+k_0 D_u  + \alpha_u + \tilde{\theta}_u V(t) \right] \\
& & \qquad  \times \left(
\beta^{uu} \left[D_u'' + \kappa_{s,u} D_u'\right]
+ \frac{\beta^{uu} }{\sigma_2} \left[D'''_u + D''_u(\kappa_{s,u}+\gamma_{s, u}) + (\kappa_{s,u}\gamma_{s, u}) D_u'  \right]
+ \beta^{uu} A_u \right. \\
&& \qquad \quad + \left.
\beta^{uv}\left[D_v'' + \kappa_{s,v} D_v'\right]
+ \frac{\beta^{uv}}{ \sigma_2}  \left[D'''_v + D''_v(\kappa_{s,v}+\gamma_{s, v}) + (\kappa_{s,v} \gamma_{s, v}) D_v'  \right]+ \beta^{uv} A_v
\right) .
\end{eqnarray*}
The equation for $A_u$, Eq. ~(\ref{Au1}), takes the form
\begin{eqnarray*}
A'_u = \frac{\phi_u}{(1-\phi_u) \sigma_2} \left[ D_u^{(4)} + D_u'''(\kappa_{s,u}+\gamma_{s, u}+\sigma_2)
    + D_u''(\kappa_{s,u}\gamma_{s, u} + \kappa_{s,u}\sigma_2+\gamma_{s, u}\sigma_2) + D_u'(\kappa_{s,u}\gamma_{s, u}\sigma_2) \right] - \mu_{u} A_u .
\end{eqnarray*}

Similarly,  the equation of $A_v$ and the high order equation for $D_v$ are:
\begin{eqnarray*}
D^{(5)}_v &=&-\left[   k_3 D_v^{(4)} + k_2 D_v''' +  k_1 D_v'' +k_0 D_v'  \right] -
\left[D_v^{(4)} + k_3 D_v''' + k_2 D_v'' + k_1D_v'+k_0 D_v  + \alpha_v - \tilde{\theta}_v V(t) \right] \\
& & \qquad  \times \left(
\beta^{vu}\left[D_u'' + \kappa_{s,u} D_u'\right]
+ \frac{\beta^{vu} }{ \sigma_2} \left[D'''_u + D''_u(\kappa_{s,u}+\gamma_{s, u}) + (\kappa_{s,u}\gamma_{s, u})
D_u'  \right]+ \beta^{vu} A_u \right. \\
&& \qquad \quad + \left.
\beta^{vv}\left[D_v'' + \kappa_{s,v} D_v'\right]
+ \frac{\beta^{vv}}{\sigma_2}  \left[D'''_v + D''_v(\kappa_{s,v}+\gamma_{s, v}) + (\kappa_{s,v}\gamma_{s, v}) D_v'  \right]+ \beta^{vv} A_v
\right) , \\
A'_v &=& \frac{\phi_v}{(1-\phi_v) \sigma_2} \left[ D_v^{(4)} + D_v'''(\kappa_{s,v}+\gamma_{s, v}+\sigma_2)
    + D_v''(\kappa_{s,v}\gamma_{s, v} + \kappa_{s,v}\sigma_2+\gamma_{s, v}\sigma_2) +
 D_v'(\kappa_{s,v}\gamma_{s, v}\sigma_2) \right] - \mu_{v} A_v .
\end{eqnarray*}

Up to now, we rewrote the original system as a system of $D_u^{(5)}, D_v^{(5)}, A_u, A_v$. It can be written as
a first-order system (by using $D, D', D'', D^{(3)}, D^{(4)}$) so that the identifiability analysis package \textit{StructuralIdentifiability}
 can be applied to find the identifiability property of the 18 parameters of this new system:
$$
\beta^{uu}, \beta^{uv}, \beta^{vu}, \beta^{vv}, \sigma_1, \sigma_2, \phi_u, \phi_v, \gamma_{s, u}, \gamma_{s, v}, \mu_u,  \mu_v,
\kappa_{s,u}, \kappa_{s,v}, \tilde{\theta}_u, \tilde{\theta}_v , \alpha_u, \alpha_v
$$
Then, the identifiability property of the 19 parameters of the original system
$$
\beta^{uu}, \beta^{uv}, \beta^{vu}, \beta^{vv}, \sigma_1, \sigma_2, \theta, \phi_u, \phi_v, \mu_u, \mu_v, \gamma_{r, u}, \gamma_{h, u},
\gamma_{r, v}, \gamma_{h, v}, \kappa_{r, u}, \kappa_{d, u}, \kappa_{r, v}, \kappa_{d, v}
$$
can be derived.
Furthermore, one may use the \textit{SIAN Webapp},
with the globally identifiable parameters (from
\textit{StructuralIdentifiability}) as extra outputs, to find the identifiability
property of the initial conditions. Identifiability results obtained from these calculations
are reported in appropriate sections in the main text.


\begin{thebibliography}{10}

\bibitem{kermack}
W.O. Kermack, A.G. McKendrick,
\newblock Contributions to the mathematical theory of epidemics --- I,
\newblock {B. Math. Biol.}, 53 (1991) 33.

\bibitem{hethcote}
H.W. Hethcote,
\newblock {SIAM Rev.}, 42 (2000) 599.

\bibitem{castillo2011}
F.~Brauer, C.~Castillo-Ch\'avez,
\newblock {Mathematical Models in Population Biology and Epidemiology},
\newblock Springer-Verlag, 2012.

\bibitem{chen2014modeling}
D. Chen,
\newblock Modeling the spread of infectious diseases: A review,
\newblock {in: Analyzing and modeling spatial and temporal dynamics of
  infectious diseases}, 2014, p. 19.

\bibitem{ForeCast}
\url{https://covid19forecasthub.org}.

\bibitem{cao21}
L.~Cao, Q.~Liu,
\newblock COVID-19 Modeling: A Review,
\newblock https://arxiv.org/abs/2104.12556.

\bibitem{shakeel}
S.M. Shakeel, N.S. Kumar, P.P. Madalli, R.~Srinivasaiah, D.R. Swamy,
\newblock Covid-19 prediction models: a systematic literature review,
\newblock {Osong Public Health Res. Perspect.}, 12 (2021) 215.

\bibitem{review_meta}
D. Calvetti, A.P. Hoover, J. Rose, E. Somersalo,
\newblock {Front. Phys.}, 8 (2020) 261.

\bibitem{bertozzi2020}
A.L. Bertozzi, E. Franco, G. Mohler, M.B. Short, D. Sledge,
\newblock {P. Natl. Acad. Sci.}, 117 (2020) 16732.

\bibitem{holmdahl2020}
I.~Holmdahl, C.~Buckee,
\newblock {New Engl. J. Med.}, 383 (2020) 303.

\bibitem{tregoning2021}
J.S. Tregoning, K.E. Flight, S.L. Higham, Z. Wang, B.F. Pierce,
\newblock {Nat. Rev. Immunol.}, 21 (2021) 626.

\bibitem{Vaccines}
\url{https://www.uptodate.com/contents/covid-19-vaccines}.

\bibitem{wagner2022}
C.E. Wagner, C.M. Saad-Roy, B.T. Grenfell,
\newblock {Nat. Rev. Immunol.}, 22 (2022) 139.

\bibitem{StilianakisVaccine2022}
G.~Angelov, R. Kovacevic, N.I. Stilianakis, V.~M. Veliov,
\newblock {Cent. Eur. J. Oper. Res.}, 31 (2022) 499.

\bibitem{marinov2022}
T.T. Marinov, R.S. Marinova,
\newblock {Sci. Rep.}, 12 (2022) 15688.

\bibitem{vacc2021}
T.~Usherwood, Z.~LaJoie, V.~Srivastava,
\newblock {Sci. Rep.}, 11 (2021) 12051.

\bibitem{MACINTYRE20222506},
C.R. MacIntyre, V. Costantino, M. Trent,
\newblock {Vaccine}, 40 (2022) 2506.

\bibitem{10.1093/cid/ciab517}
R. Kahn, I. Holmdahl, S. Reddy, J. Jernigan, M.J. Mina, R.B. Slayton,
\newblock {Clin. Infect. Dis.}, 74 (2021) 597.

\bibitem{rychtar2021}
J. Rychtar, M.L. Diagne, H. Rwezaura, S.Y. Tchoumi, J.M. Tchuenche,
\newblock {Comput. Math. Method M.} (2021) 1250129.

\bibitem{cuevas2021}
J.~Cuevas-Maraver, P.G. Kevrekidis, Q.Y. Chen, G.A. Kevrekidis, Z.~Rapti, Y.~Drossinos,
\newblock {Math. Biosci.}, 336 (2021) 108590.

\bibitem{Asymptomatic}
Q. Ma, J. Liu, Q. Liu, L. Kang, R. Liu, W. Jing, Y. Wu, M. Liu,
\newblock {JAMA Network Open}, 4 (2021) e2137257.

\bibitem{10.1093/cid/ciab916}
S.E. Waldman, T. Buehring, D.J. Escobar, S.K. Gohil, R. Gonzales, S.S. Huang, K. Olenslager, K.K. Prabaker, T. Sandoval, J. Yim, D.S. Yokoe, S.H. Cohen,
\newblock {Clin. Infect. Dis.}, 75 (2021) e895.

\bibitem{jama2022}
Y.~Wu, L. Kang, Z. Guo, J. Liu, M. Liu, W. Liang,
\newblock {JAMA Network Open}, 5 (2022) e2228008.

\bibitem{Slatere189}
T.A. Slater, S. Straw, M. Drozd, S. Kamalathasan, A. Cowley, K.K. Witte,
\newblock {Clin. Med.}, 20 (2020) e189.

\bibitem{Germany_Omicron}
\url{https://www.latimes.com/world-nation/story/2021-11-12/europe-germany-covid-hospot}.

\bibitem{TheNetherlands_Omicron}
\url{https://www.theguardian.com/world/2021/nov/30/omicron-covid-variant-present-in-europe-at-least-10-days-ago}.

\bibitem{WHO_Omicron}
\url{https://www.who.int/news/item/26-11-2021-classification-of-omicron-(b.1.1.529)-sars-cov-2-variant-of-concern}.

\bibitem{healthdata}
B.~Reiner,
\newblock Covid-19 model update: Omicron and waning immunity,
\newblock
  \url{http://www.healthdata.org/special-analysis/omicron-and-waning-immunity},
  2021.

\bibitem{Chowell}
G.~Chowell,
\newblock {Infect. Dis. Model.}, 2 (2017) 379.

\bibitem{Eisenberg2013}
J.H.~Tien, M.C.~Eisenberg, S.L.~Robertson,
\newblock {J. Theor. Biol.}, 324 (2013) 84.

\bibitem{Identify_PINN}
H. Hu, C.M. Kennedy, P.G. Kevrekidis, H.-K. Zhang,
\newblock {Viruses}, 14 (2022) 2464.

\bibitem{Pogudin_SIAN}
H. Hong, A. Ovchinnikov, G. Pogudin, C. Yap,
\newblock {Bioinformatics}, 35 (2019) 2873.

\bibitem{Pogudin_2021}
A. Ovchinnikov, A. Pillay, G. Pogudin, T. Scanlon,
\newblock {Syst. Control Lett.}, 157 (2021) 105030.

\bibitem{Pogudin_2022}
R. Dong, C. Goodbrake, H. Harrington, G. Pogudin,
\newblock Differential elimination for dynamical models via projections with
  applications to structural identifiability,
\newblock SIAM J. Appl. Algebra Geom. (2023)

\bibitem{Sloppy_2007}
R.N. Gutenkunst, J.J. Waterfall, F.P. Casey, K.S. Brown, C.R. Myers, J.P. Sethna,
\newblock {PLOS Comput. Biol.}, 3 (2007) e189

\bibitem{TimeSeriesAndalusia}
\url{https://github.com/montera34/escovid19data/blob/master/data/output/covid19-ccaa-spain_consolidated.csv.}

\bibitem{AndalusiaVaccine}
\url{https://www.juntadeandalucia.es/institutodeestadisticaycartografia/badea/informe/anual?idNode=74172}.

\bibitem{AsymptomaticsReview2022}
W. Yu, Y. Guo, S. Zhang, Y. Kong, Z. Shen, J. Zhang.
\newblock {J. Med. Virol.}, 94 (2022) 5790.

\bibitem{OpenDataSwiss}
\url{https://opendata.swiss/en/dataset/covid-19-schweiz}.

\bibitem{deathsVsHosp}
\url{https://www.vozpopuli.com/sanidad/cifras-muertos-covid.html}.

\bibitem{10.3389/fimmu.2022.812606}
N. Kumar, S. Quadri, A.I. AlAwadhi, M. AlQahtani,
\newblock {Front. Immunol.}, 13 (2022) 812606.

\bibitem{vespi2008}
V.~Colizza and A.~Vespignani,
\newblock {J. Theor. Biol.}, 251 (2008) 450.

\bibitem{rapti2022}
Z.~Rapti, J.~Cuevas-Maraver, E.~Kontou, S.~Liu, Y.~Drossinos, P.G. Kevrekidis, G.A. Kevrekidis, M.~Barmann, Q.-Y. Chen,
\newblock The role of mobility in the dynamics of the COVID-19 epidemic in {A}ndalusia,
\newblock https://arxiv.org/abs/2207.01958.

\bibitem{mammeri2020}
Y.~Mammeri,
\newblock {Comput. Math. Biophys.}, 8 (2020) 102.

\bibitem{viguerie2021}
A.~Viguerie, G.~Lorenzo, F.~Auricchio, D.~Baroli, T.J.R. Hughes, A.~Patton, A.~Reali, T.~E. Yankeelov, A.~Veneziani,
\newblock {Appl. Math. Lett.}, 111 (2021) 106617.

\bibitem{kevrekidis2021}
P.G. Kevrekidis, J.~Cuevas-Maraver, Y.~Drossinos, Z.~Rapti, G.A. Kevrekidis,
\newblock {Phys. Rev. E}, 104 (2021) 024412.

\bibitem{theo}
A.~Vaziry, T.~Kolokolnikov, P.G. Kevrekidis,
\newblock {R. Soc. Open Sci.}, 9 (2022) 220064.

\bibitem{agecovid}
V. Ram, L.P. Schaposnik,
\newblock {Sci. Rep.}, 11 (2021) 15194.

\end{thebibliography}
\end{document}